\documentclass[nofootinbib,aps,prc]{revtex4-1}

\usepackage{epsfig}
\usepackage[utf8]{inputenc}
\usepackage{amsmath}
\usepackage{amssymb}
\usepackage{amsthm}
\usepackage{amsfonts}
\usepackage{xspace}
\usepackage{graphicx}
\usepackage{graphics}
\usepackage{slashed}
\usepackage{bbold}
\usepackage{bbm}
\usepackage{physics}
\usepackage{empheq}
\usepackage{mathrsfs}
\usepackage{xcolor}

\usepackage{hyperref}

\graphicspath{ {./plots/} }

\newcommand{\Nc}{\ensuremath{N_c}\xspace}
\newcommand{\LONc}{\ensuremath{\text{LO-in-\Nc}}\xspace}

\newcommand{\NNLONc}{\ensuremath{\text{N${}^2$LO-in-\Nc}}\xspace}

\newcommand{\oneS}{{{}^{1}\!S_0}}
\newcommand{\threeS}{{{}^{3}\!S_1}}

\newcommand{\del}{\nabla}

\newcommand{\calO}{\ensuremath{\mathcal{O}}}

\newcommand{\Lagr}{\mathcal{L}}

\newcommand{\1}{\mathbbm{1}}

\newcommand{\Lone}{\ensuremath{{}^{\nopi} L_1}\xspace}
\newcommand{\Ltwo}{\ensuremath{{}^{\nopi} L_2}\xspace}

\newcommand{\CMs}{\ensuremath{C^{(M)}_s}\xspace}
\newcommand{\CMv}{\ensuremath{C^{(M)}_v}\xspace}

\newcommand{\mpi}{\ensuremath{m_\pi}}

\newcommand{\nopi}{\ensuremath{\pi\hskip-0.40em /}}
\newcommand{\eftnopi}{EFT$_{\nopi}$\xspace}
\newcommand{\LambdaNoPion}{\Lambda_{\nopi}\xspace}

\begin{document}

\title{Large-$\Nc$ analysis of magnetic and axial two-nucleon currents in pionless effective field theory}

\author{Thomas R.~Richardson}
\email{richa399@email.sc.edu}
\author{Matthias R.~Schindler}
\email{mschindl@mailbox.sc.edu}
\affiliation{Department of Physics and Astronomy,\\
University of South Carolina, \\
Columbia, SC 29208}

\date{\today}

\begin{abstract}

We analyze magnetic and axial two-nucleon contact terms in a combined large-\Nc and pionless effective field theory expansion. These terms play important roles in correctly describing, e.g., the low-energy cross section of radiative neutron capture and the deuteron magnetic moment. We show that the large-\Nc expansion hints towards a hierarchy between the two leading-order magnetic terms that matches that found in phenomenological fits. We also comment on the issue of naturalness in different Lagrangian bases. 

\end{abstract}

\maketitle

\section{Introduction}
    \label{sec:intro}

Electroweak probes of nuclear systems play an important role in testing our understanding of nucleon interactions, see., e.g., Refs.~\cite{Bacca:2014tla,Marcucci:2015rca} for recent reviews and references therein. Effective field theories (EFTs) have emerged as an important technique for studying  nucleon-nucleon (NN) interactions. 
Here we consider pionless EFT (\eftnopi), valid for energies and momenta well below the pion mass, in which only nucleons and external fields are considered as dynamical degrees of freedom. For reviews see, e.g., Refs.~\cite{Chen:1999tn,vanKolck:1999mw,Beane:2000fx,Bedaque:2002mn,Platter:2009gz} and references therein. In \eftnopi, NN interactions are described by contact operators that contain an increasing number of derivatives for increased precision. 
Electroweak fields are treated as external fields, to which nucleon operators couple based on symmetry properties. 
Two classes of contributions appear: the first class comes from minimal coupling, i.e., gauging derivatives acting on the nucleons. The second class corresponds to operators that do not originate in minimal coupling, but that are still allowed by the underlying symmetries. 
In the following, we focus on the latter, but a brief discussion of the former is included in an appendix. 
Each independent operator is multiplied by a low-energy coupling (LEC) that encodes the short-distance details of the underlying theory. For terms that couple to external fields through minimal coupling, the LECs are the same as for the corresponding terms without any external fields, and constraints on these couplings can be obtained, e.g., from scattering and bound state data. On the other hand, each independent operator in the second class introduces a new LEC, which needs to be determined from either the underlying theory (if possible, e.g., via a lattice simulation) or by comparison to experimental results where available. 
A prominent example for the second class of interactions can be found in the radiative neutron capture on protons at very low energies. A magnetic NN contact operator is required by power counting, and its inclusion agrees with the experimental result \cite{Chen:1999tn}. 
Another example is the coupling of two nucleons to an isovector axial field with an LEC $L_{1,A}$, which contributes to several important weak processes, e.g., neutrino-induced deuteron breakup, tritium $\beta$-decay, and proton-proton fusion.\footnote{In the theory with pions, chiral symmetry relates this operator to a two-nucleon-one-pion term, which also contributes to three-nucleon interactions \cite{Epelbaum:2002vt,Gardestig:2006hj}.}

In principle, the LECs should be calculable from the underlying Standard Model, which requires a nonperturbative quantum chromodynamics (QCD) calculation. Recently, the Nuclear Physics with Lattice Quantum Chromodynamics (NPLQCD) collaboration performed a first determination of $L_{1,A}$ \cite{Savage:2016kon}, though at an unphysical value of the pion mass. 
While this result holds great promise that additional calculations of LECs will be available in the future, in the meantime other theoretical constraints will be valuable, especially in cases where data may be sparse or have large uncertainties. Here we will continue previous efforts \cite{Kaplan:1995yg,Schindler:2015nga,Schindler:2018irz} to obtain such constraints by combining \eftnopi with the large-\Nc expansion \cite{tHooft:1973jz,Witten:1979kh}, where \Nc is the number of colors in QCD. 
These constraints arise from additional symmetries that emerge in QCD in the large-\Nc limit \cite{Carone:1993dz,Dashen:1993jt,Dashen:1994qi,Jenkins:1998wy,Luty:1993fu}. Previous work focused on NN interactions, both in the parity-conserving \cite{Kaplan:1995yg,Schindler:2018irz} and parity-violating \cite{Schindler:2015nga} sectors, as well as in time-reversal-invariance-violating interactions \cite{Vanasse:2019fzl}. 
In this paper we extend the application of this technique to magnetic and axial two-body currents, which, as discussed above, play important phenomenological roles. 
We find that isoscalar magnetic and axial contact terms are $N_c$-suppressed relative to their isovector counterparts. For the magnetic case, this agrees with available fits to data. In the axial case, no determination of the isoscalar term has been performed, and our results suggest that it is smaller than might be naively expected. The large-\Nc scaling of some minimal coupling terms was also considered in Ref.~\cite{Riska:2002vn}.

One of the assumptions of the EFT formalism is that LECs are expected to be ``natural,'' i.e., that the dimensionless numerical coefficients, once the appropriate scales have been factored out, should be of order 1.\footnote{Unfortunately, the term naturalness can have other meanings in different contexts.} While this is a rather vague concept and not a rigorous expectation, it has recently been used in Bayesian parameter estimations from data through the introduction of prior probability density functions (PDFs) \cite{Schindler:2008fh,Furnstahl:2015rha,Wesolowski:2015fqa,Wesolowski:2018lzj,Melendez:2019izc}. 
However, the naturalness of LECs is only a working assumption, and short-distance details of the underlying interactions could lead to apparently unnatural values for some LECs. In addition, the form of the operators entering a Lagrangian at a given order is not unique, and different functional forms can be related through, e.g., field redefinitions or Fierz identities. While observables must be independent of the formulation, large-\Nc counting is applied to LECs and hence sensitive to the formulation. 
We refer to a given set of operators defining a Lagrangian as a basis and will show in an example below that naturalness in one basis can be hidden in a different basis that is physically equivalent.

In Sec.~\ref{sec:methods} we present aspects of \eftnopi and the large-\Nc expansion relevant for our analysis. Magnetic two-nucleon contact terms are considered in Sec.~\ref{sec:magnetic} and the generalization to axial currents is described in Sec.~\ref{sec:axial}. The large-\Nc scaling of terms originating from minimal coupling in the two-derivative Lagrangian is discussed in the Appendix.

\section{Pionless effective field theory and the large-\Nc expansion}
    \label{sec:methods}

Pionless EFT describes the interactions of nucleons with each other and with external fields through a series of contact operators. Each operator is accompanied by a low-energy coupling (LEC) that encodes all short-distance/high-energy physics. Since the pion has been integrated out as an active degree of freedom, \eftnopi is valid only at energies well below the pion mass. Observables are expanded in terms of the small ratio $Q/\LambdaNoPion$, with $Q$ the energy/momentum transfer, while $\LambdaNoPion \sim \mpi$ is the breakdown scale of the theory. 

The determination of the relative importance of terms in an EFT Lagrangian is called power counting. Power counting in \eftnopi is not simply based on the number of derivatives acting on an operator \cite{Kaplan:1996xu,Kaplan:1998we,Kaplan:1998tg,vanKolck:1998bw}. A consistent power counting can be achieved by use of the power divergent subtraction (PDS) scheme \cite{Kaplan:1998tg}, which introduces an additional scale into the theory through the renormalization scale $\mu$. The scale $\mu$ needs to be on the order of the scale $Q$ defined above for the power counting to be valid. While physical quantities do not depend on the renormalization scale, LECs are not observables and in general are $\mu$-dependent.

In the following we combine \eftnopi with the large-\Nc expansion.
The basic ingredient in applying the large-\Nc analysis in the context of two-nucleon interactions is the observation that the baryon matrix elements of different combinations of spin-isospin operators have different large-\Nc scalings \cite{Dashen:1994qi,Kaplan:1995yg}. This is the approach used in the analyses of two-nucleon forces in Refs.~\cite{Kaplan:1995yg, Kaplan:1996rk}, and Ref.~\cite{Phillips:2013rsa} contains a review of this method and an extension to three-nucleon forces. 
In the context of \eftnopi, it has been shown \cite{Kaplan:1995yg,Schindler:2015nga,Schindler:2018irz} that in the large-\Nc limit additional constraints on the relative size of various LECs contributing to nucleon-nucleon scattering can be derived. 
Here, we extend the analysis to two-nucleon contact terms that couple to a magnetic, and more generally an axial, external field. 
The relevant quantity to consider is the matrix element of the Hamiltonian between states that contain two nucleons and, for the case considered below, an additional external field $A$ in one of the states,
\begin{equation}
    \bra{N_\gamma N_\delta A} H \ket{N_\alpha N_\beta} ,
\end{equation}
where $\alpha,\beta,\gamma, \delta$ denote combined spin and isospin quantum numbers, and the external field can also carry spin and isospin quantum numbers. 
For our purposes, the nucleon momenta will be irrelevant and they are kept implicit in this notation.
We assume that, as in the case without an external field, the baryonic part of the Hamiltonian takes a Hartree form and can be expanded as \cite{Witten:1979kh,Kaplan:1996rk}
\begin{equation}
\label{H:matrixelement}
    H_{\text{baryon}} = \Nc \sum_{n} \sum_{s,t} v_{stn} \left( \frac{ \hat{S}^i }{N_c} \right)^s \left( \frac{ \hat{I}^a }{N_c} \right)^t \left( \frac{ \hat{G}^{ia} }{N_c} \right)^{n-s-t} \, ,
\end{equation}
in terms of the operators
\begin{equation}
    \hat{S}^i = q^\dagger \frac{\sigma^i}{2} q, \quad \hat{I}^a = q^\dagger \frac{\tau^a}{2} q, \quad \hat{G}^{ia} = q^\dagger \frac{\sigma^i \tau^a}{4} q \, ,
\end{equation}
where the $q$ are colorless bosonic quark fields. The \Nc quarks in the ground state nucleon are totally antisymmetric in their color indices. Therefore, because the quarks are fermions, the quark fields must be completely symmetric in the spin-flavor indices \cite{Dashen:1994qi}. This permits the omission of the $q$ color index and treating the quark operator as a bosonic operator.
The coefficients $v_{stn}$ are chosen to match the required symmetry properties and in general can depend on the nucleon momenta. This Hamiltonian is then combined with the external field $A$ such that spin and/or isospin indices are appropriately contracted.

The baryonic matrix elements of Eq.~\eqref{H:matrixelement} factorize in the large-$N_c$ limit \cite{Kaplan:1995yg},
\begin{equation}
    \bra{N_\gamma N_\delta} \mathcal{O}_1 \mathcal{O}_2 \ket{N_\alpha N_\beta} \xrightarrow[]{N_c \to \infty} \bra{N_\gamma} \mathcal O_1 \ket{N_\alpha} \bra{N_\delta} \mathcal O_2 \ket{N_\beta} + \text{crossed},
\end{equation}
and the large-\Nc scaling of single-baryon matrix elements is given by \cite{Dashen:1993jt,Dashen:1994qi,Kaplan:1995yg,Kaplan:1996rk}
\begin{equation}
    \bra{N_\gamma} \frac{\calO^{(n)}_{I,S}}{\Nc^n} \ket{N_\alpha} \lesssim \frac{1}{\Nc^{\vert I-S \vert}}\, ,
\end{equation}
for an $n$-body operator $\calO^{(n)}_{I,S}$ with spin $S$ and isospin $I$. In particular,
\begin{equation}
    \label{single:scaling}
    \begin{split}
    \bra{N_\gamma} \frac{\1}{\Nc} \ket{N_\alpha} &\sim \bra{N_\gamma} \frac{G^{ia}}{\Nc} \ket{N_\alpha} \lesssim 1\, ,\\
    \bra{N_\gamma} \frac{S^i}{\Nc} \ket{N_\alpha} &\sim \bra{N_\gamma} \frac{I^a}{\Nc} \ket{N_\alpha} \lesssim \frac{1}{\Nc} \, .
    \end{split}
\end{equation}
The matrix elements of the Hartree Hamiltonian are then related to the corresponding matrix elements in \eftnopi by matching the spin-isospin structure of the nucleon operators to that in the quark operator expansion, and the scaling is mapped onto a large-\Nc scaling of the \eftnopi LECs. For example, the \eftnopi operator $ N^\dagger \sigma^i \tau^a N $ has the same structure as $ G^{ia} $ in the quark operator expansion, so we consider $\bra{N_\gamma} N^\dagger \sigma^i \tau^a N \ket{N_\alpha} \sim O(\Nc)$.

As in some previous work in the two-nucleon sector \cite{Kaplan:1995yg,Kaplan:1996rk,Schindler:2015nga,Schindler:2018irz}, the effects of virtual $\Delta$ degrees of freedom in the matrix elements are ignored. The $\Delta$ becomes degenerate with the nucleon in the large-\Nc limit and is known to play an important role in the meson-exchange picture of two-nucleon interactions \cite{Banerjee:2001js}. It is an open question how to properly treat $\Delta$ degrees of freedom in the combined \eftnopi and large-\Nc expansion, but previous work \cite{Kaplan:1995yg,Kaplan:1996rk,Schindler:2018irz} indicates that not considering their effects explicitly can still lead to results that agree with experimental information. $\Delta$ intermediate states in NN scattering were discussed in Ref.~\cite{Savage:1996tb}.

\section{Magnetic two-nucleon contact terms}
    \label{sec:magnetic}

At leading order (LO) in the \eftnopi expansion, there are two independent two-nucleon contact terms coupling to the magnetic field $\vb B$.
In the partial-wave basis, these can be written as \cite{Chen:1999tn}
\begin{equation}
        \label{external:EM:lagr_pw}
    \Lagr= e B_i \left[ \Lone \left( N^T P_i N \right)^\dagger \left( N^T \bar{P}_3 N \right) - i \epsilon^{ijk} \, \Ltwo \left( N^T P_j N \right)^\dagger \left( N^T P_k N \right)\right] +\text{h.c.},
\end{equation}
where $P_i = \frac{1}{\sqrt{8}} \sigma_2\sigma_i \tau_2$ and $\bar{P}_a = \frac{1}{\sqrt{8}} \sigma_2 \tau_2\tau_a $ are the projection operators onto the $\threeS$ and $\oneS$ partial waves, respectively, with $\sigma_i$ ($\tau_a$) the Pauli matrices acting in spin (isospin) space. Alternatively, the Lagrangian can be expressed in a different basis, in which the large-\Nc counting rules can be made manifest. In this basis, a minimal form is given by \cite{Pastore:2009is}
\begin{equation}
    \Lagr = e B_i \left\{C_{15}^\prime \left( N^\dagger \sigma_i N \right) \left( N^\dagger N \right) + C_{16}^\prime \left[\left( N^\dagger \sigma^i \tau^3 N \right) \left( N^\dagger N \right) - \left( N^\dagger \sigma^i  N \right) \left( N^\dagger \tau^3 N \right) \right] \right\}.
\end{equation}
The minimal form of the Lagrangian is not unique; it is obtained by applying Fierz transformations to relate various terms that are consistent with the required symmetry properties \cite{Pastore:2009is}, but there is freedom in choosing which terms to retain. An important consideration in making this choice is that the application of Fierz transformations can hide the correct large-\Nc scaling of operators \cite{Schindler:2015nga}.

To address this issue, we first consider the most general set of two-nucleon operators coupled to an external magnetic field, analyze the large-\Nc scaling of each term, and then retain that  scaling while using Fierz transformations to construct a minimal operator set. Power counting and parity invariance  dictate that at LO in the \eftnopi only operators without derivatives acting on the nucleon fields are present. The two-nucleon operators must be $U(1)$ gauge invariant and Hermitian, and must transform as a vector under rotations. This interaction is not restricted to isoscalars, and may include nontrivial isospin structure. The Lagrangian can be written as
    \begin{align}
            \label{external:EM:nonminimal}
        \Lagr & = e B^i \left[ \tilde{C}^{(M)}_1 \left( N^\dagger \sigma^i N \right) \left( N^\dagger N \right) + \tilde{C}^{(M)}_2 \left( N^\dagger \sigma^i \tau^a N \right) \left( N^\dagger \tau^a N \right) \right. \nonumber \\
        &\quad + \tilde{C}^{(M)}_3 \epsilon^{ijk} \epsilon^{3ab} \left( N^\dagger \sigma^j \tau^a N \right) \left( N^\dagger \sigma^k \tau^b N \right) +\tilde{C}^{(M)}_4  \left( N^\dagger \sigma^i \tau^3 N \right) \left( N^\dagger N \right) \\
        & \quad \left. + \tilde{C}^{(M)}_5  \left( N^\dagger \sigma^i N \right) \left( N^\dagger \tau^3 N \right) \right].  \nonumber
    \end{align}
The large-\Nc counting rules of Eq.~\eqref{single:scaling} lead to the following scaling of the LECs:
\begin{align}
    \tilde{C}^{(M)}_1 &\sim O(\Nc^0)\, , \\
    \tilde{C}^{(M)}_2 &\sim O(\Nc^0)\, , \\
    \tilde{C}^{(M)}_3 &\sim O(\Nc)\, , \\
    \tilde{C}^{(M)}_4 &\sim O(\Nc)\, , \\
    \tilde{C}^{(M)}_5 &\sim O(\Nc^{-1})\, .
\end{align}
While it seems that there are two operators that are of LO in the large-\Nc counting (\LONc), the operators in Eq.~\eqref{external:EM:nonminimal} are not independent, and the application of Fierz transformations reduces the Lagrangian to two independent operators.  While in principle any minimal set of independent operators is equivalent, we choose to retain the terms with the manifestly dominant large-\Nc scaling in each isospin sector. This yields an isovector term that differs from that used in Ref.~\cite{Pastore:2009is}, and the Lagrangian takes the form 
    \begin{equation}
            \label{external:EM:min_lagr}
        \Lagr = e B^i \left[ \CMs \left( N^\dagger \sigma^i N \right) \left( N^\dagger N \right)
        +\CMv \epsilon^{ijk} \epsilon^{3ab} \left( N^\dagger \sigma^j \tau^a N \right) \left( N^\dagger \sigma^k \tau^b N \right) \right],
    \end{equation}
where
    \begin{alignat}{2}
            \label{external:EM:nonmin_scaling}
        \CMs &= \tilde C^{(M)}_1 - 3 \tilde C^{(M)}_2 &&\sim O(\Nc^0) \, , \\
        \CMv &= \tilde C^{(M)}_3 + \frac{1}{4} \left( \tilde C^{(M)}_4 - \tilde C^{(M)}_5 \right) &&\sim O(\Nc) \, . 
    \end{alignat}
We thus find that there is only a single independent \LONc term, with the isovector coupling to the magnetic field $\sim$ $O(\Nc)$, while the isoscalar coupling is suppressed and $\sim O(\Nc^0)$.
    
A different set of Fierz transformations relates these LECs to those in the partial-wave basis of Eq.~\eqref{external:EM:lagr_pw},
    \begin{equation}
            \label{external:EM:Chen_relation}
        \Lone = 8 \CMv , \quad \Ltwo = - \CMs,
    \end{equation}
which means that \Ltwo is suppressed by $1/\Nc$ compared to \Lone. This suppression is  mentioned in Ref.~\cite{Detmold:2004qn} for operators in the dibaryon formalism.

The LECs appearing in the effective Lagrangian are, absent any additional constraints, often assumed to be natural, i.e., of $O(1)$ after extracting appropriate dimensionful parameters. While the definition of what this means is not exact, it implies that two LECs appearing at the same order should be approximately of the same size.
$\Lone$ can be obtained from a fit to the experimental cross section for the radiative capture $np \rightarrow d \gamma$, while $\Ltwo$ can be determined from the deuteron magnetic moment. The values of these LECs are renormalization-scale dependent, and at the scale $\mu = m_\pi$ are given by \cite{Chen:1999tn}
\begin{equation}
\label{eq:mag_exp_values}
        \Lone = 7.24 \, \text{fm}^4 \, , \quad \Ltwo = -0.149 \, \text{fm}^4 \, .
\end{equation}
It is argued in Ref.~\cite{Chen:1999vd} that the value of $\Ltwo$ ``is significantly smaller than the naively estimated size of $\sim 1 \, \text{fm}^4$.'' 
The ratio of the two fitted LECs at $\mu=m_\pi$ is
    \begin{equation}
            \label{external:EM:ratio}
        \abs{ \frac{ \Ltwo }{ \Lone } }_{\text{exp}} \approx 0.021 \, ,
    \end{equation}
which clearly shows that it is challenging to consider both \Lone and \Ltwo to be simultaneously natural at $\mu=\mpi$.

The large-\Nc analysis above shows that the isoscalar contribution is suppressed compared to the isovector one by a factor of \Nc, with $\Nc = 3$ in the real world. While this provides a hint for why the isoscalar contribution is smaller, it cannot account for the factor of 50. However, the transition through Fierz transformations from the large-\Nc basis to the partial-wave basis in which \Lone and \Ltwo are defined introduces an additional suppression factor of 1/8. 
Thus, if instead of \Lone and \Ltwo we assume that $\CMs$ and $\CMv$ are of the same size aside from the large-\Nc suppression, i.e., $\vert \CMv \vert \sim \Nc \vert \CMs \vert$, then the predicted ratio at the physical value $\Nc=3$ is given by
    \begin{equation}
            \label{external:EM:constraint}
        \abs{ \frac{\Ltwo}{\Lone}}_{\Nc} \approx \frac{1}{8 N_c} \approx 0.042 \, .
    \end{equation}
Using Eq.~\eqref{external:EM:Chen_relation} and the values for \Lone and \Ltwo of Eq.~\eqref{eq:mag_exp_values} gives (at $\mu=\mpi$)
\begin{equation}
\label{eq:mag-C-num}
    \CMs = 0.149 \, \text{fm}^4 , \quad \CMv = 0.905 \, \text{fm}^4 \, ,
\end{equation}
i.e., $\CMv$ is larger than $\CMs$ by a factor of $\approx 6$. Taking into account the large-\Nc suppression, the remaining difference in the two values is about a factor of 2, which is easily accommodated within naturalness.  In other words, aside from the large-\Nc suppression, $\CMs$ and $\CMv$ can be considered of the same size, i.e., $\vert \CMv \vert \sim \Nc \vert \CMs \vert$.
Since the large-$N_c$ constraints are really upper bounds on the LECs, it appears that the large-\Nc constraint is not in disagreement with the values of the coefficients obtained in Ref.~\cite{Chen:1999tn}.

As previously mentioned, the transition from the large-\Nc basis to the partial wave basis can introduce additional factors of 8 due to the normalization of projection operators.
These factors could in principle be avoided by defining the projection operators differently. However, the projection operators $P^{(s)}$ for a given partial wave $s$ are defined to satisfy \cite{Fleming:1999ee}
\begin{equation}
    \sum_\text{pol. avg} \text{Tr}\left[ P^{(s)} P^{(s)\dagger} \right] = \frac{1}{2}.
\end{equation}
Since they also appear in other two-nucleon operators, e.g., those contributing to NN scattering, changing the normalization in one case would introduce corresponding factors in the LECs of other operators.

This result seems to indicate that the naturalness of LECs might be hidden in one basis, while it is more apparent in others.
On the other hand, implementation of the PDS renormalization scheme is more straightforward in the partial-wave basis, making it more practical.\footnote{For issues connected to Fierz transformations relating various forms of short-range three-nucleon interactions, see Refs.~\cite{Lynn:2015jua,Lynn:2017fxg,Lonardoni:2018nob}.}

Of course, this example does not allow us to draw any general conclusions about how suitable a given basis is for assuming natural LECs. 
However, it does indicate that special care should be taken when trying to quantify naturalness, for example when incorporating the naturalness assumption into any functional form of prior PDFs on the LECs. 

\begin{figure}
        \centering
        \includegraphics[scale = 0.7]{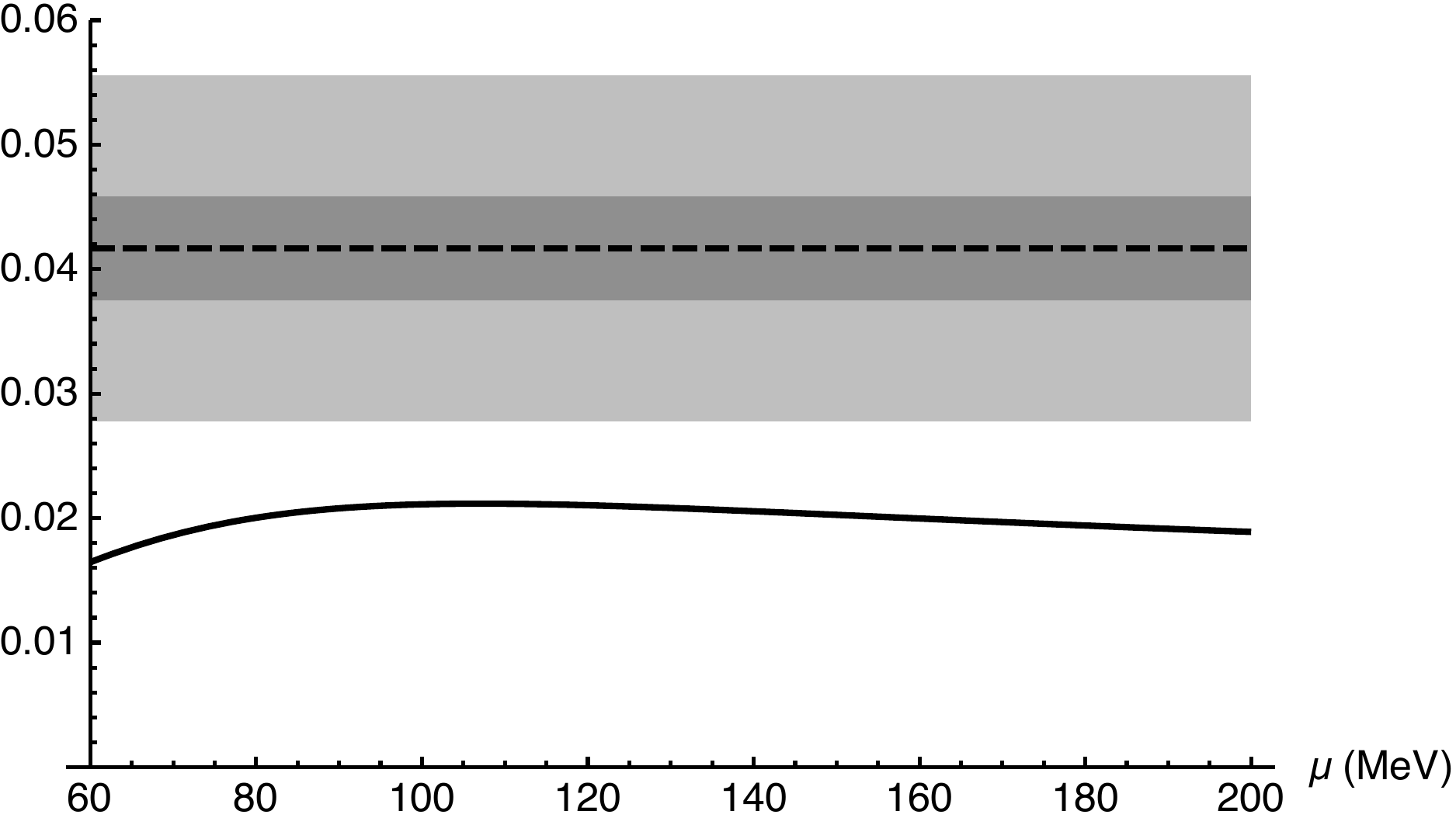}
        \caption{The ratio $\vert\Ltwo / \Lone \vert$ as a function of the renormalization scale $\mu$. The solid line corresponds to solutions of the renormalization group equations combined with the values of Eq.~\eqref{eq:mag_exp_values}. The dashed line corresponds to Eq.~\eqref{external:EM:constraint}, while the dark (light) gray band corresponds to 10\% (30\%) corrections.}
        \label{fig:L1L2}
    \end{figure}

Both LECs \Lone and \Ltwo are renormalization scale dependent and the values used so far correspond to $\mu=\mpi$ in the PDS scheme. As discussed in Refs.~\cite{Kaplan:1995yg,Schindler:2018irz}, the large-\Nc relationships can be hidden if a different choice is made for the subtraction point $\mu$. The results of Ref.~\cite{Schindler:2018irz} indicate that $\mu\approx \mpi$ tends to be within the region where the large-\Nc constraints on the LECs  do not contradict the data. For completeness, Fig.~\ref{fig:L1L2} shows the running of the ratio $\vert \Ltwo / \Lone \vert$ with $\mu$. 
The ratio is relatively flat starting around $80\, \text{MeV}$ and no significant differences to the values at $\mu=\mpi$ are seen.

\section{Axial two-nucleon contact terms}
    \label{sec:axial}

The analysis of the previous section can be generalized to the coupling of two-nucleon operators to an arbitrary external axial-vector field. Again, the terms at LO in \eftnopi contain no derivatives. The isoscalar sector proceeds in complete analogy to the magnetic case, with the magnetic LECs replaced by axial ones, $\tilde{C}^{(M)}_1 \to \tilde{C}^{(A)}_1, \tilde{C}^{(M)}_2 \to \tilde{C}^{(A)}_2$. The same Fierz relationships hold and there is a single independent isoscalar term, which we again choose to be
\begin{equation}
    \Lagr_A^s = A^i C^{(A)}_s \left( N^\dagger \sigma^i N \right) \left( N^\dagger N \right),
\end{equation}
with 
\begin{equation}
    C^{(A)}_s = \tilde C^{(A)}_1 - 3 \tilde C^{(A)}_2 \sim O(\Nc^0) .
\end{equation} In the partial-wave basis, this operator again takes the form
\begin{equation}
   \Lagr_A^s = - 2 i \epsilon^{ijk} L_{2,A} A^i \left( N^T P_j N \right)^\dagger \left( N^T P_k N \right),
\end{equation}
with $L_{2,A} = - C^{(A)}_s$.

For the coupling to an isovector axial-vector field $A^{ia}$, the nonminimal set again contains three operators as in the magnetic case, with the only change the replacement of the $z$ component of isospin by a general isospin index $a$ and the change to axial LECs, $\tilde{C}^{(M)}_i \to \tilde{C}^{(A)}_i \ (i=3,4,5)$. Analogous Fierz identities hold and only one term is linearly independent. We choose the manifestly large-\Nc dominant term
\begin{equation}
    \Lagr_A^v = A^{ia} C^{(A)}_v \epsilon^{ijk} \epsilon^{abc} \left( N^\dagger \sigma^j \tau^b N \right) \left( N^\dagger \sigma^k \tau^c N \right),
\end{equation}
with 
\begin{equation}
    C^{(A)}_v = \tilde{C}^{(A)}_3 +\frac{1}{4}\left(  \tilde C^{(A)}_4 - \tilde C^{(A)}_5 \right) \sim O(\Nc).
\end{equation}
The partial-wave expression for this Lagrangian is \cite{Butler:1999sv}
\begin{equation}
    \Lagr_A^v = A^{ia} L_{1,A} \left( N^T P_i N \right)^\dagger \left( N^T \bar{P}_a N \right)+\text{h.c.}.
\end{equation}
Application of Fierz transformations shows that
    \begin{equation}
        L_{1,A} = 8 C^{(A)}_v,
    \end{equation}
in complete analogy to the magnetic case. Therefore, the same prediction for the ratio of LECs as for the magnetic LECs holds,
    \begin{equation}
    \label{external:axial:ratio}
        \left \vert \frac{L_{2,A}}{L_{1,A}} \right \vert_{\Nc}  = \left \vert \frac{C^{(A)}_s}{8 C^{(A)}_v} \right \vert \sim \frac{1}{8 \Nc},
    \end{equation}
where for the last relation we again assume that the only relative suppression between the isoscalar and isovector LECs comes from the large-\Nc scaling, i.e., $ \vert C^{(A)}_v \vert \sim \Nc \vert C^{(A)}_s\vert $.

From na\"ive dimensional analysis (NDA), the values of these LECs are estimated to be \cite{Butler:1999sv}
    \begin{equation}
        \vert L_{1,A}\vert \sim \vert L_{2,A} \vert \sim \frac{4\pi}{M} \frac{1}{\mu^2} \approx 5 \, \text{fm}^3,
    \end{equation}
 at $\mu=\mpi$, and in this approach are both expected to be of the same size. There have been several efforts to constrain the value of $L_{1,A}$ from proton-proton fusion, neutrino-deuteron reactions, and tritium $\beta$-decay \cite{Butler:1999sv,Butler:2002cw,Chen:2002pv,Ando:2008va,De-Leon:2016wyu,Acharya:2019fij}. 
The NPLQCD collaboration recently performed a lattice QCD calculation of the proton-proton fusion process at $m_\pi^\text{lat} = 806 \, \text{MeV}$ and used the result to extract the LEC $L_{1,A}$ with $\mu$ at the physical pion mass, obtaining \cite{Savage:2016kon} 
    \begin{equation}
        L_{1,A}^{\text{NPLQCD}} (\mu=m_\pi) = 3.9(0.2)(1.0)(0.4)(0.9) \, \text{fm}^3,
    \end{equation}
where the values in parentheses denote the statistical, systematic fitting and analysis, and systematic mass extrapolation uncertainties, as well as an estimate of higher-order corrections in the \eftnopi power counting.
Within uncertainties, this value is in agreement with the extractions based on neutrino-deuteron reactions and tritium $\beta$ decay.
These studies do not attempt to determine the value of $L_{2,A}$ since its contribution to the processes of interest is negligible. 

While in NDA the magnitudes of $L_{1,A}$ and $L_{2,A}$ are expected to be of the same size, the results for the magnetic LECs suggest that it might be useful to assume naturalness in the large-\Nc basis combined with the large-\Nc suppression of the isoscalar coupling, i.e., assuming that $\vert C^{(A)}_v \vert \approx \Nc \vert C^{(A)}_s \vert$. 
Using, for example, the NPLQCD value in the large-$N_c$ relationship of Eq.~\eqref{external:axial:ratio} with $\Nc=3$ would result in
    \begin{equation}
        L_{2,A} \sim 0.1625 \, \text{fm}^3\, ,
    \end{equation}
which would provide even stronger justification for neglecting this contribution in the processes that have been considered.

\section{Conclusions}

We analyzed two-nucleon contact operators coupling to magnetic and general axial external fields in a combined \eftnopi and large-\Nc expansion. These operators provide important contributions to a variety of physical processes, such as low-energy radiative neutron capture, tritium $\beta$-decay, and proton-proton fusion.
In \eftnopi, the LECs accompanying the operators are treated as parameters that are expected to be of natural size. The large-\Nc analysis, on the other hand, shows an additional hierarchy, with the isoscalar coupling suppressed by a factor of 1/\Nc relative to the isovector coupling. 

In the magnetic case, the isoscalar (\Ltwo) and isovector (\Lone) LECs have been determined from fits to the deuteron magnetic moment and the $np\to d\gamma$ cross section, respectively. Their ratio, evaluated at the renormalization scale $\mu=\mpi$, is approximately 1/50, which is far outside the naive large-\Nc expectation of 1/3 and raises the concern whether or not both LECs can be considered natural.
However, an additional factor of 8 enters the Fierz  transformation from the operator basis in which the large-\Nc analysis is performed to the basis in which \Lone and \Ltwo are defined. We showed that taking into account the 1/\Nc suppression, the residual values of the LECs in this large-Nc basis can be of approximately the same size and still be consistent with the small ratio in terms of the partial-wave LECs. 
This may indicate that the naturalness of LECs can be hidden in some choices for the set of operators describing the same physics. Special care should thus be taken when trying to quantify naturalness, e.g., in defining Bayesian priors.

Only the isovector axial LEC $L_{1,A}$ has been determined so far. Assuming the validity of the analysis in the magnetic case, the isoscalar axial LEC $L_{2,A}$ may be significantly smaller than naively expected. However, it is important to not treat the large-\Nc scalings as exact predictions, but rather as indicators of general trends.

\begin{acknowledgments}
This material is based upon work supported by the U.S. Department of Energy, Office of Science, Office of Nuclear Physics, under Award Number DE-SC0019647. We thank S.~Pastore and R.~P.~Springer for interesting and helpful discussions, as well as comments on the manuscript.
\end{acknowledgments}

\appendix*

\section{Minimal Coupling for Two-Derivative Contact Interactions}
    \label{MinimalCoup}

We make use of the usual prescription for the covariant derivative with minimal coupling,
    \begin{equation}
        \mathcal D^i N = \del^{i}N - ie A^{i} Q N,
    \end{equation}
where $Q$ is the charge matrix $ Q = \frac{1}{2} \left( \1 + \tau^3 \right) $. The derivatives for the two-derivative operators in Ref.~\cite{Schindler:2018irz} may be replaced with the covariant derivative to produce gauge invariant interactions. This approach was first adopted for the two-derivative contact terms in Ref.~\cite{Pastore:2008ui}, but we choose a different operator basis in the following. When the derivative replacement is made, there will be two-nucleon-one-photon and two-nucleon-two-photon interactions. Although they are paramount to maintaining gauge invariance, the two-photon interactions will not be considered. If the bilinears involved in the interactions are of the form
    \begin{equation}
        \mathcal D \left( N^\dagger \calO_1 N \right) \mathcal D \left( N^\dagger \calO_2 N \right),
    \end{equation}
where $ \calO_1 $ and $ \calO_2 $ only contain Pauli spin matrices, the operator does not couple to the electromagnetic field.
Therefore, we will not need to consider the terms proportional to $ C_{1 \cdot 1}, \ C_{\sigma \cdot \sigma}, \ \text{and } C'_{\sigma \cdot \sigma} $ in the notation of Ref.~\cite{Schindler:2018irz}.

The current operators can be obtained through
    \begin{equation}
        J^\mu = \partial_\nu \frac{\partial \Lagr}{\partial(\partial_\nu A_\mu)} - \frac{\partial \Lagr}{\partial A_\mu}
    \end{equation}
in a manner similar to Ref. \cite{Kolling:2011mt}. At \LONc, the current operators are
    \begin{align}
        J^i_{G \cdot G} & =  2 e C_{G \cdot G} \epsilon^{3ab} \del^i(N^\dagger \sigma^j \tau^a N) (N^\dagger \sigma^j \tau^b N) \, ,    \\
        J^{^\prime i}_{G \cdot G} & =  2 e C^\prime_{G \cdot G} \epsilon^{3ab} \del^j(N^\dagger \sigma^j \tau^a N) (N^\dagger \sigma^i \tau^b N)   \, ,
    \end{align}
 while at next-to-next-to-leading order in the large-\Nc scaling (\NNLONc) they are
\begin{align}
    \overset\leftrightarrow{J}^i_{1 \cdot 1} & =  4 i e \overset\leftrightarrow{C}_{1 \cdot 1} (N^\dagger \overset\leftrightarrow{\del}^i N)(N^\dagger Q N) \, , \\
        J^i_{\tau \cdot \tau} & =  2 e C_{\tau \cdot \tau} \epsilon^{3ab} \del^i (N^\dagger  \tau^a N) (N^\dagger  \tau^b N) \, ,   \\
    \overset\leftrightarrow{J}^i_{G \cdot G} & =  2ie \overset\leftrightarrow{C}_{G \cdot G} \left[ (N^\dagger \sigma^j \tau^a \overset\leftrightarrow{\del}^i  N) (N^\dagger \sigma^j \tau^a   N)  + (N^\dagger \sigma^j \tau^3 \overset\leftrightarrow{\del}^i  N) (N^\dagger \sigma^j   N) \right]  \, ,  \\
    \overset\leftrightarrow{J}^i_{1 \cdot \sigma} & =   e \overset\leftrightarrow{C}_{1 \cdot \sigma} \epsilon^{ijk} \left[           \del^k (N^\dagger \sigma^j N)(N^\dagger Q N) + \del^k (N^\dagger N) (N^\dagger \sigma^j Q N) \right]   \, ,  \\
    \overset\leftrightarrow{J^\prime}^i_{G \cdot G} & =  2 i e \overset\leftrightarrow{C^\prime}_{G \cdot G} \left[ (N^\dagger \overset\leftrightarrow{\del}^j \sigma^j \tau^a N) (N^\dagger \sigma^i \tau^a N) + (N^\dagger \overset\leftrightarrow{\del}^j     \sigma^j \tau^3 N) (N^\dagger \sigma^i N) \right]  \, ,   \\
    \overset\leftrightarrow{J}^i_{G \cdot \tau} & =   \frac{e}{2} \overset\leftrightarrow{C}_{G \cdot \tau} \epsilon^{ijk} \bigg\{     \left[ \del^k \left( N^\dagger \sigma^j \tau^a N \right) \left( N^\dagger \tau^a N \right) + \del^k \left( N^\dagger          \tau^a N \right) \left( N^\dagger \sigma^j \tau^a N \right) \right. \nonumber \\
        &  \left. \left. + \del^k \left( N^\dagger \sigma^j \tau^3 N \right) \left( N^\dagger N \right) + \del^k \left( N^\dagger \tau^3 N \right) \left( N^\dagger \sigma^j N \right) \right] \right. \\
        &  + i \epsilon^{3ab} \left[ \left( N^\dagger \sigma^j \tau^b N \right) \left( N^\dagger \overset\leftrightarrow{\del}^k \tau^a N \right) + \left( N^\dagger \tau^b N \right) \left( N^\dagger \overset\leftrightarrow{\del}^k \sigma^j \tau^a N \right) \right] \bigg\} \, . \nonumber  
    \end{align}

Making use of Fierz transformations for the \NNLONc terms reduces the set of operators to
   \begin{align}
        J^i_\LONc & = 2 e C_{G \cdot G}  \epsilon^{3ab} \del^i \left( N^\dagger \sigma^j \tau^a N \right) \left( N^\dagger \sigma^j \tau^b N \right) \nonumber \\
        & \quad + 2 e C^\prime_{G \cdot G}  \epsilon^{3ab} \del^j(N^\dagger \sigma^j \tau^a N) (N^\dagger \sigma^i \tau^b N) \, , \label{Current:LONc} \\
        J^i_\NNLONc & = 2e \left( C_{\tau \cdot \tau}-\frac{1}{4}\overset\leftrightarrow{C}_{1\cdot 1} -\frac{3}{4}\overset\leftrightarrow{C}_{G\cdot G} +\frac{1}{4}\overset\leftrightarrow{C^\prime}_{G\cdot G} \right)  \epsilon^{3ab} \del^i (N^\dagger  \tau^a N) (N^\dagger  \tau^b N) \nonumber \\
        & \quad +  e \left( \overset\leftrightarrow{C}_{1 \cdot \sigma}+\overset\leftrightarrow{C}_{G \cdot \tau} \right) \epsilon^{ijk} \left[ \del^k (N^\dagger \sigma^j N)(N^\dagger Q N) + \del^k (N^\dagger N) (N^\dagger \sigma^j Q N) \right]  \,  , \label{Current:NNLONc}
    \end{align}
where we are not showing subleading corrections in $J^i_\LONc$ that enter through the Fierz transformations.
This is the same result one would obtain from first finding the minimal set of operators and then gauging them, i.e., once the dominant large-\Nc scaling has been determined, the order of gauging and performing Fierz transformations is irrelevant.

The form of these operators can be matched to the minimal-coupling contact currents in Ref.~\cite{Piarulli:2012bn}, which are equivalent to those of Ref.~\cite{Pastore:2008ui}. This shows that the second and third terms in Eq.~(2.20) of Ref.~\cite{Piarulli:2012bn} are \LONc, while the first and fourth terms are \NNLONc. When matching the LECs of Ref.~\cite{Schindler:2018irz} to those used in Ref.~\cite{Piarulli:2012bn}, $C_2$, $C_4$, and $C_7$ receive \LONc contributions, which naively suggests three \LONc terms. However, the \LONc contributions cancel in the linear combination $C_2+3C_4+C_7$, leading to the first term of Eq.~(2.20) in Ref.~\cite{Piarulli:2012bn} contributing at \NNLONc.

In Ref. \cite{Riska:2002vn}, the two-nucleon isovector electromagnetic current equivalent to the minimally-coupled terms is derived in terms of Fermi invariants, and the large-\Nc scalings are determined from the spin-flavor structure of the current. The operators and scalings  match those of Eqs.~\eqref{Current:LONc} and \eqref{Current:NNLONc}.


\begin{thebibliography}{51}%
\makeatletter
\providecommand \@ifxundefined [1]{%
 \@ifx{#1\undefined}
}%
\providecommand \@ifnum [1]{%
 \ifnum #1\expandafter \@firstoftwo
 \else \expandafter \@secondoftwo
 \fi
}%
\providecommand \@ifx [1]{%
 \ifx #1\expandafter \@firstoftwo
 \else \expandafter \@secondoftwo
 \fi
}%
\providecommand \natexlab [1]{#1}%
\providecommand \enquote  [1]{``#1''}%
\providecommand \bibnamefont  [1]{#1}%
\providecommand \bibfnamefont [1]{#1}%
\providecommand \citenamefont [1]{#1}%
\providecommand \href@noop [0]{\@secondoftwo}%
\providecommand \href [0]{\begingroup \@sanitize@url \@href}%
\providecommand \@href[1]{\@@startlink{#1}\@@href}%
\providecommand \@@href[1]{\endgroup#1\@@endlink}%
\providecommand \@sanitize@url [0]{\catcode `\\12\catcode `\$12\catcode
  `\&12\catcode `\#12\catcode `\^12\catcode `\_12\catcode `\%12\relax}%
\providecommand \@@startlink[1]{}%
\providecommand \@@endlink[0]{}%
\providecommand \url  [0]{\begingroup\@sanitize@url \@url }%
\providecommand \@url [1]{\endgroup\@href {#1}{\urlprefix }}%
\providecommand \urlprefix  [0]{URL }%
\providecommand \Eprint [0]{\href }%
\providecommand \doibase [0]{http://dx.doi.org/}%
\providecommand \selectlanguage [0]{\@gobble}%
\providecommand \bibinfo  [0]{\@secondoftwo}%
\providecommand \bibfield  [0]{\@secondoftwo}%
\providecommand \translation [1]{[#1]}%
\providecommand \BibitemOpen [0]{}%
\providecommand \bibitemStop [0]{}%
\providecommand \bibitemNoStop [0]{.\EOS\space}%
\providecommand \EOS [0]{\spacefactor3000\relax}%
\providecommand \BibitemShut  [1]{\csname bibitem#1\endcsname}%
\let\auto@bib@innerbib\@empty
\bibitem [{\citenamefont {Bacca}\ and\ \citenamefont
  {Pastore}(2014)}]{Bacca:2014tla}%
  \BibitemOpen
  \bibfield  {author} {\bibinfo {author} {\bibfnamefont {S.}~\bibnamefont
  {Bacca}}\ and\ \bibinfo {author} {\bibfnamefont {S.}~\bibnamefont
  {Pastore}},\ }\href {\doibase 10.1088/0954-3899/41/12/123002} {\bibfield
  {journal} {\bibinfo  {journal} {J. Phys.}\ }\textbf {\bibinfo {volume}
  {G41}},\ \bibinfo {pages} {123002} (\bibinfo {year} {2014})},\ \Eprint
  {http://arxiv.org/abs/1407.3490} {arXiv:1407.3490 [nucl-th]} \BibitemShut
  {NoStop}%
\bibitem [{\citenamefont {Marcucci}\ \emph {et~al.}(2016)\citenamefont
  {Marcucci}, \citenamefont {Gross}, \citenamefont {Pena}, \citenamefont
  {Piarulli}, \citenamefont {Schiavilla}, \citenamefont {Sick}, \citenamefont
  {Stadler}, \citenamefont {Van~Orden},\ and\ \citenamefont
  {Viviani}}]{Marcucci:2015rca}%
  \BibitemOpen
  \bibfield  {author} {\bibinfo {author} {\bibfnamefont {L.~E.}\ \bibnamefont
  {Marcucci}}, \bibinfo {author} {\bibfnamefont {F.}~\bibnamefont {Gross}},
  \bibinfo {author} {\bibfnamefont {M.~T.}\ \bibnamefont {Pena}}, \bibinfo
  {author} {\bibfnamefont {M.}~\bibnamefont {Piarulli}}, \bibinfo {author}
  {\bibfnamefont {R.}~\bibnamefont {Schiavilla}}, \bibinfo {author}
  {\bibfnamefont {I.}~\bibnamefont {Sick}}, \bibinfo {author} {\bibfnamefont
  {A.}~\bibnamefont {Stadler}}, \bibinfo {author} {\bibfnamefont {J.~W.}\
  \bibnamefont {Van~Orden}}, \ and\ \bibinfo {author} {\bibfnamefont
  {M.}~\bibnamefont {Viviani}},\ }\href {\doibase
  10.1088/0954-3899/43/2/023002} {\bibfield  {journal} {\bibinfo  {journal} {J.
  Phys.}\ }\textbf {\bibinfo {volume} {G43}},\ \bibinfo {pages} {023002}
  (\bibinfo {year} {2016})},\ \Eprint {http://arxiv.org/abs/1504.05063}
  {arXiv:1504.05063 [nucl-th]} \BibitemShut {NoStop}%
\bibitem [{\citenamefont {Chen}\ \emph
  {et~al.}(1999{\natexlab{a}})\citenamefont {Chen}, \citenamefont {Rupak},\
  and\ \citenamefont {Savage}}]{Chen:1999tn}%
  \BibitemOpen
  \bibfield  {author} {\bibinfo {author} {\bibfnamefont {J.-W.}\ \bibnamefont
  {Chen}}, \bibinfo {author} {\bibfnamefont {G.}~\bibnamefont {Rupak}}, \ and\
  \bibinfo {author} {\bibfnamefont {M.~J.}\ \bibnamefont {Savage}},\ }\href
  {\doibase 10.1016/S0375-9474(99)00298-5} {\bibfield  {journal} {\bibinfo
  {journal} {Nucl.Phys.}\ }\textbf {\bibinfo {volume} {A653}},\ \bibinfo
  {pages} {386} (\bibinfo {year} {1999}{\natexlab{a}})},\ \Eprint
  {http://arxiv.org/abs/nucl-th/9902056} {arXiv:nucl-th/9902056 [nucl-th]}
  \BibitemShut {NoStop}%
\bibitem [{\citenamefont {van Kolck}(1999{\natexlab{a}})}]{vanKolck:1999mw}%
  \BibitemOpen
  \bibfield  {author} {\bibinfo {author} {\bibfnamefont {U.}~\bibnamefont {van
  Kolck}},\ }\href {\doibase 10.1016/S0146-6410(99)00097-6} {\bibfield
  {journal} {\bibinfo  {journal} {Prog.Part.Nucl.Phys.}\ }\textbf {\bibinfo
  {volume} {43}},\ \bibinfo {pages} {337} (\bibinfo {year}
  {1999}{\natexlab{a}})},\ \Eprint {http://arxiv.org/abs/nucl-th/9902015}
  {arXiv:nucl-th/9902015 [nucl-th]} \BibitemShut {NoStop}%
\bibitem [{\citenamefont {Beane}\ \emph {et~al.}(2001)\citenamefont {Beane},
  \citenamefont {Bedaque}, \citenamefont {Haxton}, \citenamefont {Phillips},\
  and\ \citenamefont {Savage}}]{Beane:2000fx}%
  \BibitemOpen
  \bibfield  {author} {\bibinfo {author} {\bibfnamefont {S.~R.}\ \bibnamefont
  {Beane}}, \bibinfo {author} {\bibfnamefont {P.~F.}\ \bibnamefont {Bedaque}},
  \bibinfo {author} {\bibfnamefont {W.~C.}\ \bibnamefont {Haxton}}, \bibinfo
  {author} {\bibfnamefont {D.~R.}\ \bibnamefont {Phillips}}, \ and\ \bibinfo
  {author} {\bibfnamefont {M.~J.}\ \bibnamefont {Savage}},\ }in\ \href
  {\doibase 10.1142/9789812810458_0011} {\emph {\bibinfo {booktitle} {At The
  Frontier of Particle Physics: Handbook Of QCD}}},\ Vol.~\bibinfo {volume}
  {1},\ \bibinfo {editor} {edited by\ \bibinfo {editor} {\bibfnamefont
  {M.}~\bibnamefont {Shifman}}}\ (\bibinfo  {publisher} {World Scientific,
  Singapore},\ \bibinfo {year} {2001})\ pp.\ \bibinfo {pages} {133--269},\
  \Eprint {http://arxiv.org/abs/nucl-th/0008064} {arXiv:nucl-th/0008064
  [nucl-th]} \BibitemShut {NoStop}%
\bibitem [{\citenamefont {Bedaque}\ and\ \citenamefont {van
  Kolck}(2002)}]{Bedaque:2002mn}%
  \BibitemOpen
  \bibfield  {author} {\bibinfo {author} {\bibfnamefont {P.~F.}\ \bibnamefont
  {Bedaque}}\ and\ \bibinfo {author} {\bibfnamefont {U.}~\bibnamefont {van
  Kolck}},\ }\href {\doibase 10.1146/annurev.nucl.52.050102.090637} {\bibfield
  {journal} {\bibinfo  {journal} {Annu. Rev. Nucl. Part. Sci.}\ }\textbf
  {\bibinfo {volume} {52}},\ \bibinfo {pages} {339} (\bibinfo {year} {2002})},\
  \Eprint {http://arxiv.org/abs/nucl-th/0203055} {arXiv:nucl-th/0203055}
  \BibitemShut {NoStop}%
\bibitem [{\citenamefont {Platter}(2009)}]{Platter:2009gz}%
  \BibitemOpen
  \bibfield  {author} {\bibinfo {author} {\bibfnamefont {L.}~\bibnamefont
  {Platter}},\ }\href {\doibase 10.1007/s00601-009-0057-0} {\bibfield
  {journal} {\bibinfo  {journal} {Few Body Syst.}\ }\textbf {\bibinfo {volume}
  {46}},\ \bibinfo {pages} {139} (\bibinfo {year} {2009})},\ \Eprint
  {http://arxiv.org/abs/0904.2227} {arXiv:0904.2227 [nucl-th]} \BibitemShut
  {NoStop}%
\bibitem [{\citenamefont {Epelbaum}\ \emph {et~al.}(2002)\citenamefont
  {Epelbaum}, \citenamefont {Nogga}, \citenamefont {Gloeckle}, \citenamefont
  {Kamada}, \citenamefont {Mei{\ss}ner} \emph {et~al.}}]{Epelbaum:2002vt}%
  \BibitemOpen
  \bibfield  {author} {\bibinfo {author} {\bibfnamefont {E.}~\bibnamefont
  {Epelbaum}}, \bibinfo {author} {\bibfnamefont {A.}~\bibnamefont {Nogga}},
  \bibinfo {author} {\bibfnamefont {W.}~\bibnamefont {Gloeckle}}, \bibinfo
  {author} {\bibfnamefont {H.}~\bibnamefont {Kamada}}, \bibinfo {author}
  {\bibfnamefont {U.-G.}\ \bibnamefont {Mei{\ss}ner}},  \emph {et~al.},\ }\href
  {\doibase 10.1103/PhysRevC.66.064001} {\bibfield  {journal} {\bibinfo
  {journal} {Phys.Rev.}\ }\textbf {\bibinfo {volume} {C66}},\ \bibinfo {pages}
  {064001} (\bibinfo {year} {2002})},\ \Eprint
  {http://arxiv.org/abs/nucl-th/0208023} {arXiv:nucl-th/0208023 [nucl-th]}
  \BibitemShut {NoStop}%
\bibitem [{\citenamefont {Gardestig}\ and\ \citenamefont
  {Phillips}(2006)}]{Gardestig:2006hj}%
  \BibitemOpen
  \bibfield  {author} {\bibinfo {author} {\bibfnamefont {A.}~\bibnamefont
  {Gardestig}}\ and\ \bibinfo {author} {\bibfnamefont {D.~R.}\ \bibnamefont
  {Phillips}},\ }\href {\doibase 10.1103/PhysRevLett.96.232301} {\bibfield
  {journal} {\bibinfo  {journal} {Phys. Rev. Lett.}\ }\textbf {\bibinfo
  {volume} {96}},\ \bibinfo {pages} {232301} (\bibinfo {year} {2006})},\
  \Eprint {http://arxiv.org/abs/nucl-th/0603045} {arXiv:nucl-th/0603045
  [nucl-th]} \BibitemShut {NoStop}%
\bibitem [{\citenamefont {Savage}\ \emph {et~al.}(2017)\citenamefont {Savage},
  \citenamefont {Shanahan}, \citenamefont {Tiburzi}, \citenamefont {Wagman},
  \citenamefont {Winter}, \citenamefont {Beane}, \citenamefont {Chang},
  \citenamefont {Davoudi}, \citenamefont {Detmold},\ and\ \citenamefont
  {Orginos}}]{Savage:2016kon}%
  \BibitemOpen
  \bibfield  {author} {\bibinfo {author} {\bibfnamefont {M.~J.}\ \bibnamefont
  {Savage}}, \bibinfo {author} {\bibfnamefont {P.~E.}\ \bibnamefont
  {Shanahan}}, \bibinfo {author} {\bibfnamefont {B.~C.}\ \bibnamefont
  {Tiburzi}}, \bibinfo {author} {\bibfnamefont {M.~L.}\ \bibnamefont {Wagman}},
  \bibinfo {author} {\bibfnamefont {F.}~\bibnamefont {Winter}}, \bibinfo
  {author} {\bibfnamefont {S.~R.}\ \bibnamefont {Beane}}, \bibinfo {author}
  {\bibfnamefont {E.}~\bibnamefont {Chang}}, \bibinfo {author} {\bibfnamefont
  {Z.}~\bibnamefont {Davoudi}}, \bibinfo {author} {\bibfnamefont
  {W.}~\bibnamefont {Detmold}}, \ and\ \bibinfo {author} {\bibfnamefont
  {K.}~\bibnamefont {Orginos}},\ }\href {\doibase
  10.1103/PhysRevLett.119.062002} {\bibfield  {journal} {\bibinfo  {journal}
  {Phys. Rev. Lett.}\ }\textbf {\bibinfo {volume} {119}},\ \bibinfo {pages}
  {062002} (\bibinfo {year} {2017})},\ \Eprint
  {http://arxiv.org/abs/1610.04545} {arXiv:1610.04545 [hep-lat]} \BibitemShut
  {NoStop}%
\bibitem [{\citenamefont {Kaplan}\ and\ \citenamefont
  {Savage}(1996)}]{Kaplan:1995yg}%
  \BibitemOpen
  \bibfield  {author} {\bibinfo {author} {\bibfnamefont {D.~B.}\ \bibnamefont
  {Kaplan}}\ and\ \bibinfo {author} {\bibfnamefont {M.~J.}\ \bibnamefont
  {Savage}},\ }\href {\doibase 10.1016/0370-2693(95)01277-X} {\bibfield
  {journal} {\bibinfo  {journal} {Phys. Lett.}\ }\textbf {\bibinfo {volume}
  {B365}},\ \bibinfo {pages} {244} (\bibinfo {year} {1996})},\ \Eprint
  {http://arxiv.org/abs/hep-ph/9509371} {arXiv:hep-ph/9509371 [hep-ph]}
  \BibitemShut {NoStop}%
\bibitem [{\citenamefont {Schindler}\ \emph {et~al.}(2016)\citenamefont
  {Schindler}, \citenamefont {Springer},\ and\ \citenamefont
  {Vanasse}}]{Schindler:2015nga}%
  \BibitemOpen
  \bibfield  {author} {\bibinfo {author} {\bibfnamefont {M.~R.}\ \bibnamefont
  {Schindler}}, \bibinfo {author} {\bibfnamefont {R.~P.}\ \bibnamefont
  {Springer}}, \ and\ \bibinfo {author} {\bibfnamefont {J.}~\bibnamefont
  {Vanasse}},\ }\href {\doibase 10.1103/PhysRevC.97.059901,
  10.1103/PhysRevC.93.025502} {\bibfield  {journal} {\bibinfo  {journal} {Phys.
  Rev.}\ }\textbf {\bibinfo {volume} {C93}},\ \bibinfo {pages} {025502}
  (\bibinfo {year} {2016})},\ \bibinfo {note} {[Erratum: Phys.
  Rev.C97,no.5,059901(2018)]},\ \Eprint {http://arxiv.org/abs/1510.07598}
  {arXiv:1510.07598 [nucl-th]} \BibitemShut {NoStop}%
\bibitem [{\citenamefont {Schindler}\ \emph {et~al.}(2018)\citenamefont
  {Schindler}, \citenamefont {Singh},\ and\ \citenamefont
  {Springer}}]{Schindler:2018irz}%
  \BibitemOpen
  \bibfield  {author} {\bibinfo {author} {\bibfnamefont {M.~R.}\ \bibnamefont
  {Schindler}}, \bibinfo {author} {\bibfnamefont {H.}~\bibnamefont {Singh}}, \
  and\ \bibinfo {author} {\bibfnamefont {R.~P.}\ \bibnamefont {Springer}},\
  }\href {\doibase 10.1103/PhysRevC.98.044001} {\bibfield  {journal} {\bibinfo
  {journal} {Phys. Rev.}\ }\textbf {\bibinfo {volume} {C98}},\ \bibinfo {pages}
  {044001} (\bibinfo {year} {2018})},\ \Eprint
  {http://arxiv.org/abs/1805.06056} {arXiv:1805.06056 [nucl-th]} \BibitemShut
  {NoStop}%
\bibitem [{\citenamefont {'t~Hooft}(1974)}]{tHooft:1973jz}%
  \BibitemOpen
  \bibfield  {author} {\bibinfo {author} {\bibfnamefont {G.}~\bibnamefont
  {'t~Hooft}},\ }\href@noop {} {\bibfield  {journal} {\bibinfo  {journal}
  {Nucl. Phys.}\ }\textbf {\bibinfo {volume} {B72}},\ \bibinfo {pages} {461}
  (\bibinfo {year} {1974})}\BibitemShut {NoStop}%
\bibitem [{\citenamefont {Witten}(1979)}]{Witten:1979kh}%
  \BibitemOpen
  \bibfield  {author} {\bibinfo {author} {\bibfnamefont {E.}~\bibnamefont
  {Witten}},\ }\href {\doibase 10.1016/0550-3213(79)90232-3} {\bibfield
  {journal} {\bibinfo  {journal} {Nucl. Phys.}\ }\textbf {\bibinfo {volume}
  {B160}},\ \bibinfo {pages} {57} (\bibinfo {year} {1979})}\BibitemShut
  {NoStop}%
\bibitem [{\citenamefont {Carone}\ \emph {et~al.}(1994)\citenamefont {Carone},
  \citenamefont {Georgi},\ and\ \citenamefont {Osofsky}}]{Carone:1993dz}%
  \BibitemOpen
  \bibfield  {author} {\bibinfo {author} {\bibfnamefont {C.}~\bibnamefont
  {Carone}}, \bibinfo {author} {\bibfnamefont {H.}~\bibnamefont {Georgi}}, \
  and\ \bibinfo {author} {\bibfnamefont {S.}~\bibnamefont {Osofsky}},\ }\href
  {\doibase 10.1016/0370-2693(94)91112-6} {\bibfield  {journal} {\bibinfo
  {journal} {Phys. Lett.}\ }\textbf {\bibinfo {volume} {B322}},\ \bibinfo
  {pages} {227} (\bibinfo {year} {1994})},\ \Eprint
  {http://arxiv.org/abs/hep-ph/9310365} {arXiv:hep-ph/9310365 [hep-ph]}
  \BibitemShut {NoStop}%
\bibitem [{\citenamefont {Dashen}\ \emph {et~al.}(1994)\citenamefont {Dashen},
  \citenamefont {Jenkins},\ and\ \citenamefont {Manohar}}]{Dashen:1993jt}%
  \BibitemOpen
  \bibfield  {author} {\bibinfo {author} {\bibfnamefont {R.~F.}\ \bibnamefont
  {Dashen}}, \bibinfo {author} {\bibfnamefont {E.~E.}\ \bibnamefont {Jenkins}},
  \ and\ \bibinfo {author} {\bibfnamefont {A.~V.}\ \bibnamefont {Manohar}},\
  }\href {\doibase 10.1103/PhysRevD.51.2489, 10.1103/PhysRevD.49.4713}
  {\bibfield  {journal} {\bibinfo  {journal} {Phys. Rev.}\ }\textbf {\bibinfo
  {volume} {D49}},\ \bibinfo {pages} {4713} (\bibinfo {year} {1994})},\
  \bibinfo {note} {[Erratum: Phys. Rev.D51,2489(1995)]},\ \Eprint
  {http://arxiv.org/abs/hep-ph/9310379} {arXiv:hep-ph/9310379 [hep-ph]}
  \BibitemShut {NoStop}%
\bibitem [{\citenamefont {Dashen}\ \emph {et~al.}(1995)\citenamefont {Dashen},
  \citenamefont {Jenkins},\ and\ \citenamefont {Manohar}}]{Dashen:1994qi}%
  \BibitemOpen
  \bibfield  {author} {\bibinfo {author} {\bibfnamefont {R.~F.}\ \bibnamefont
  {Dashen}}, \bibinfo {author} {\bibfnamefont {E.~E.}\ \bibnamefont {Jenkins}},
  \ and\ \bibinfo {author} {\bibfnamefont {A.~V.}\ \bibnamefont {Manohar}},\
  }\href {\doibase 10.1103/PhysRevD.51.3697} {\bibfield  {journal} {\bibinfo
  {journal} {Phys. Rev.}\ }\textbf {\bibinfo {volume} {D51}},\ \bibinfo {pages}
  {3697} (\bibinfo {year} {1995})},\ \Eprint
  {http://arxiv.org/abs/hep-ph/9411234} {arXiv:hep-ph/9411234 [hep-ph]}
  \BibitemShut {NoStop}%
\bibitem [{\citenamefont {Jenkins}(1998)}]{Jenkins:1998wy}%
  \BibitemOpen
  \bibfield  {author} {\bibinfo {author} {\bibfnamefont {E.~E.}\ \bibnamefont
  {Jenkins}},\ }\href {\doibase 10.1146/annurev.nucl.48.1.81} {\bibfield
  {journal} {\bibinfo  {journal} {Ann. Rev. Nucl. Part. Sci.}\ }\textbf
  {\bibinfo {volume} {48}},\ \bibinfo {pages} {81} (\bibinfo {year} {1998})},\
  \Eprint {http://arxiv.org/abs/hep-ph/9803349} {arXiv:hep-ph/9803349 [hep-ph]}
  \BibitemShut {NoStop}%
\bibitem [{\citenamefont {Luty}\ and\ \citenamefont
  {March-Russell}(1994)}]{Luty:1993fu}%
  \BibitemOpen
  \bibfield  {author} {\bibinfo {author} {\bibfnamefont {M.~A.}\ \bibnamefont
  {Luty}}\ and\ \bibinfo {author} {\bibfnamefont {J.}~\bibnamefont
  {March-Russell}},\ }\href {\doibase 10.1016/0550-3213(94)90126-0} {\bibfield
  {journal} {\bibinfo  {journal} {Nucl. Phys.}\ }\textbf {\bibinfo {volume}
  {B426}},\ \bibinfo {pages} {71} (\bibinfo {year} {1994})},\ \Eprint
  {http://arxiv.org/abs/hep-ph/9310369} {arXiv:hep-ph/9310369 [hep-ph]}
  \BibitemShut {NoStop}%
\bibitem [{\citenamefont {Vanasse}\ and\ \citenamefont
  {David}(2019)}]{Vanasse:2019fzl}%
  \BibitemOpen
  \bibfield  {author} {\bibinfo {author} {\bibfnamefont {J.}~\bibnamefont
  {Vanasse}}\ and\ \bibinfo {author} {\bibfnamefont {A.}~\bibnamefont
  {David}},\ }\href@noop {} {\  (\bibinfo {year} {2019})},\ \Eprint
  {http://arxiv.org/abs/1910.03133} {arXiv:1910.03133 [nucl-th]} \BibitemShut
  {NoStop}%
\bibitem [{\citenamefont {Riska}(2002)}]{Riska:2002vn}%
  \BibitemOpen
  \bibfield  {author} {\bibinfo {author} {\bibfnamefont {D.~O.}\ \bibnamefont
  {Riska}},\ }\href {\doibase 10.1016/S0375-9474(02)01091-6} {\bibfield
  {journal} {\bibinfo  {journal} {Nucl. Phys.}\ }\textbf {\bibinfo {volume}
  {A710}},\ \bibinfo {pages} {55} (\bibinfo {year} {2002})},\ \Eprint
  {http://arxiv.org/abs/nucl-th/0204016} {arXiv:nucl-th/0204016 [nucl-th]}
  \BibitemShut {NoStop}%
\bibitem [{\citenamefont {Schindler}\ and\ \citenamefont
  {Phillips}(2009)}]{Schindler:2008fh}%
  \BibitemOpen
  \bibfield  {author} {\bibinfo {author} {\bibfnamefont {M.~R.}\ \bibnamefont
  {Schindler}}\ and\ \bibinfo {author} {\bibfnamefont {D.~R.}\ \bibnamefont
  {Phillips}},\ }\href {\doibase 10.1016/j.aop.2008.09.003,
  10.1016/j.aop.2009.05.007} {\bibfield  {journal} {\bibinfo  {journal} {Annals
  Phys.}\ }\textbf {\bibinfo {volume} {324}},\ \bibinfo {pages} {682} (\bibinfo
  {year} {2009})},\ \bibinfo {note} {[Erratum: Annals Phys.324,2051(2009)]},\
  \Eprint {http://arxiv.org/abs/0808.3643} {arXiv:0808.3643 [hep-ph]}
  \BibitemShut {NoStop}%
\bibitem [{\citenamefont {Furnstahl}\ \emph {et~al.}(2015)\citenamefont
  {Furnstahl}, \citenamefont {Klco}, \citenamefont {Phillips},\ and\
  \citenamefont {Wesolowski}}]{Furnstahl:2015rha}%
  \BibitemOpen
  \bibfield  {author} {\bibinfo {author} {\bibfnamefont {R.~J.}\ \bibnamefont
  {Furnstahl}}, \bibinfo {author} {\bibfnamefont {N.}~\bibnamefont {Klco}},
  \bibinfo {author} {\bibfnamefont {D.~R.}\ \bibnamefont {Phillips}}, \ and\
  \bibinfo {author} {\bibfnamefont {S.}~\bibnamefont {Wesolowski}},\ }\href
  {\doibase 10.1103/PhysRevC.92.024005} {\bibfield  {journal} {\bibinfo
  {journal} {Phys. Rev.}\ }\textbf {\bibinfo {volume} {C92}},\ \bibinfo {pages}
  {024005} (\bibinfo {year} {2015})},\ \Eprint
  {http://arxiv.org/abs/1506.01343} {arXiv:1506.01343 [nucl-th]} \BibitemShut
  {NoStop}%
\bibitem [{\citenamefont {Wesolowski}\ \emph {et~al.}(2016)\citenamefont
  {Wesolowski}, \citenamefont {Klco}, \citenamefont {Furnstahl}, \citenamefont
  {Phillips},\ and\ \citenamefont {Thapaliya}}]{Wesolowski:2015fqa}%
  \BibitemOpen
  \bibfield  {author} {\bibinfo {author} {\bibfnamefont {S.}~\bibnamefont
  {Wesolowski}}, \bibinfo {author} {\bibfnamefont {N.}~\bibnamefont {Klco}},
  \bibinfo {author} {\bibfnamefont {R.~J.}\ \bibnamefont {Furnstahl}}, \bibinfo
  {author} {\bibfnamefont {D.~R.}\ \bibnamefont {Phillips}}, \ and\ \bibinfo
  {author} {\bibfnamefont {A.}~\bibnamefont {Thapaliya}},\ }\href {\doibase
  10.1088/0954-3899/43/7/074001} {\bibfield  {journal} {\bibinfo  {journal} {J.
  Phys.}\ }\textbf {\bibinfo {volume} {G43}},\ \bibinfo {pages} {074001}
  (\bibinfo {year} {2016})},\ \Eprint {http://arxiv.org/abs/1511.03618}
  {arXiv:1511.03618 [nucl-th]} \BibitemShut {NoStop}%
\bibitem [{\citenamefont {Wesolowski}\ \emph {et~al.}(2019)\citenamefont
  {Wesolowski}, \citenamefont {Furnstahl}, \citenamefont {Melendez},\ and\
  \citenamefont {Phillips}}]{Wesolowski:2018lzj}%
  \BibitemOpen
  \bibfield  {author} {\bibinfo {author} {\bibfnamefont {S.}~\bibnamefont
  {Wesolowski}}, \bibinfo {author} {\bibfnamefont {R.~J.}\ \bibnamefont
  {Furnstahl}}, \bibinfo {author} {\bibfnamefont {J.~A.}\ \bibnamefont
  {Melendez}}, \ and\ \bibinfo {author} {\bibfnamefont {D.~R.}\ \bibnamefont
  {Phillips}},\ }\href {\doibase 10.1088/1361-6471/aaf5fc} {\bibfield
  {journal} {\bibinfo  {journal} {J. Phys.}\ }\textbf {\bibinfo {volume}
  {G46}},\ \bibinfo {pages} {045102} (\bibinfo {year} {2019})},\ \Eprint
  {http://arxiv.org/abs/1808.08211} {arXiv:1808.08211 [nucl-th]} \BibitemShut
  {NoStop}%
\bibitem [{\citenamefont {Melendez}\ \emph {et~al.}(2019)\citenamefont
  {Melendez}, \citenamefont {Furnstahl}, \citenamefont {Phillips},
  \citenamefont {Pratola},\ and\ \citenamefont
  {Wesolowski}}]{Melendez:2019izc}%
  \BibitemOpen
  \bibfield  {author} {\bibinfo {author} {\bibfnamefont {J.~A.}\ \bibnamefont
  {Melendez}}, \bibinfo {author} {\bibfnamefont {R.~J.}\ \bibnamefont
  {Furnstahl}}, \bibinfo {author} {\bibfnamefont {D.~R.}\ \bibnamefont
  {Phillips}}, \bibinfo {author} {\bibfnamefont {M.~T.}\ \bibnamefont
  {Pratola}}, \ and\ \bibinfo {author} {\bibfnamefont {S.}~\bibnamefont
  {Wesolowski}},\ }\href {\doibase 10.1103/PhysRevC.100.044001} {\bibfield
  {journal} {\bibinfo  {journal} {Phys. Rev.}\ }\textbf {\bibinfo {volume}
  {C100}},\ \bibinfo {pages} {044001} (\bibinfo {year} {2019})},\ \Eprint
  {http://arxiv.org/abs/1904.10581} {arXiv:1904.10581 [nucl-th]} \BibitemShut
  {NoStop}%
\bibitem [{\citenamefont {Kaplan}\ \emph {et~al.}(1996)\citenamefont {Kaplan},
  \citenamefont {Savage},\ and\ \citenamefont {Wise}}]{Kaplan:1996xu}%
  \BibitemOpen
  \bibfield  {author} {\bibinfo {author} {\bibfnamefont {D.~B.}\ \bibnamefont
  {Kaplan}}, \bibinfo {author} {\bibfnamefont {M.~J.}\ \bibnamefont {Savage}},
  \ and\ \bibinfo {author} {\bibfnamefont {M.~B.}\ \bibnamefont {Wise}},\
  }\href {\doibase 10.1016/0550-3213(96)00357-4} {\bibfield  {journal}
  {\bibinfo  {journal} {Nucl.Phys.}\ }\textbf {\bibinfo {volume} {B478}},\
  \bibinfo {pages} {629} (\bibinfo {year} {1996})},\ \Eprint
  {http://arxiv.org/abs/nucl-th/9605002} {arXiv:nucl-th/9605002 [nucl-th]}
  \BibitemShut {NoStop}%
\bibitem [{\citenamefont {Kaplan}\ \emph
  {et~al.}(1998{\natexlab{a}})\citenamefont {Kaplan}, \citenamefont {Savage},\
  and\ \citenamefont {Wise}}]{Kaplan:1998we}%
  \BibitemOpen
  \bibfield  {author} {\bibinfo {author} {\bibfnamefont {D.~B.}\ \bibnamefont
  {Kaplan}}, \bibinfo {author} {\bibfnamefont {M.~J.}\ \bibnamefont {Savage}},
  \ and\ \bibinfo {author} {\bibfnamefont {M.~B.}\ \bibnamefont {Wise}},\
  }\href {\doibase 10.1016/S0550-3213(98)00440-4} {\bibfield  {journal}
  {\bibinfo  {journal} {Nucl.Phys.}\ }\textbf {\bibinfo {volume} {B534}},\
  \bibinfo {pages} {329} (\bibinfo {year} {1998}{\natexlab{a}})},\ \Eprint
  {http://arxiv.org/abs/nucl-th/9802075} {arXiv:nucl-th/9802075 [nucl-th]}
  \BibitemShut {NoStop}%
\bibitem [{\citenamefont {Kaplan}\ \emph
  {et~al.}(1998{\natexlab{b}})\citenamefont {Kaplan}, \citenamefont {Savage},\
  and\ \citenamefont {Wise}}]{Kaplan:1998tg}%
  \BibitemOpen
  \bibfield  {author} {\bibinfo {author} {\bibfnamefont {D.~B.}\ \bibnamefont
  {Kaplan}}, \bibinfo {author} {\bibfnamefont {M.~J.}\ \bibnamefont {Savage}},
  \ and\ \bibinfo {author} {\bibfnamefont {M.~B.}\ \bibnamefont {Wise}},\
  }\href {\doibase 10.1016/S0370-2693(98)00210-X} {\bibfield  {journal}
  {\bibinfo  {journal} {Phys.Lett.}\ }\textbf {\bibinfo {volume} {B424}},\
  \bibinfo {pages} {390} (\bibinfo {year} {1998}{\natexlab{b}})},\ \Eprint
  {http://arxiv.org/abs/nucl-th/9801034} {arXiv:nucl-th/9801034 [nucl-th]}
  \BibitemShut {NoStop}%
\bibitem [{\citenamefont {van Kolck}(1999{\natexlab{b}})}]{vanKolck:1998bw}%
  \BibitemOpen
  \bibfield  {author} {\bibinfo {author} {\bibfnamefont {U.}~\bibnamefont {van
  Kolck}},\ }\href {\doibase 10.1016/S0375-9474(98)00612-5} {\bibfield
  {journal} {\bibinfo  {journal} {Nucl. Phys.}\ }\textbf {\bibinfo {volume}
  {A645}},\ \bibinfo {pages} {273} (\bibinfo {year} {1999}{\natexlab{b}})},\
  \Eprint {http://arxiv.org/abs/nucl-th/9808007} {arXiv:nucl-th/9808007
  [nucl-th]} \BibitemShut {NoStop}%
\bibitem [{\citenamefont {Kaplan}\ and\ \citenamefont
  {Manohar}(1997)}]{Kaplan:1996rk}%
  \BibitemOpen
  \bibfield  {author} {\bibinfo {author} {\bibfnamefont {D.~B.}\ \bibnamefont
  {Kaplan}}\ and\ \bibinfo {author} {\bibfnamefont {A.~V.}\ \bibnamefont
  {Manohar}},\ }\href {\doibase 10.1103/PhysRevC.56.76} {\bibfield  {journal}
  {\bibinfo  {journal} {Phys.Rev.}\ }\textbf {\bibinfo {volume} {C56}},\
  \bibinfo {pages} {76} (\bibinfo {year} {1997})},\ \Eprint
  {http://arxiv.org/abs/nucl-th/9612021} {arXiv:nucl-th/9612021 [nucl-th]}
  \BibitemShut {NoStop}%
\bibitem [{\citenamefont {Phillips}\ and\ \citenamefont
  {Schat}(2013)}]{Phillips:2013rsa}%
  \BibitemOpen
  \bibfield  {author} {\bibinfo {author} {\bibfnamefont {D.~R.}\ \bibnamefont
  {Phillips}}\ and\ \bibinfo {author} {\bibfnamefont {C.}~\bibnamefont
  {Schat}},\ }\href {\doibase 10.1103/PhysRevC.88.034002} {\bibfield  {journal}
  {\bibinfo  {journal} {Phys. Rev.}\ }\textbf {\bibinfo {volume} {C88}},\
  \bibinfo {pages} {034002} (\bibinfo {year} {2013})},\ \Eprint
  {http://arxiv.org/abs/1307.6274} {arXiv:1307.6274 [nucl-th]} \BibitemShut
  {NoStop}%
\bibitem [{\citenamefont {Banerjee}\ \emph {et~al.}(2002)\citenamefont
  {Banerjee}, \citenamefont {Cohen},\ and\ \citenamefont
  {Gelman}}]{Banerjee:2001js}%
  \BibitemOpen
  \bibfield  {author} {\bibinfo {author} {\bibfnamefont {M.~K.}\ \bibnamefont
  {Banerjee}}, \bibinfo {author} {\bibfnamefont {T.~D.}\ \bibnamefont {Cohen}},
  \ and\ \bibinfo {author} {\bibfnamefont {B.~A.}\ \bibnamefont {Gelman}},\
  }\href {\doibase 10.1103/PhysRevC.65.034011} {\bibfield  {journal} {\bibinfo
  {journal} {Phys. Rev.}\ }\textbf {\bibinfo {volume} {C65}},\ \bibinfo {pages}
  {034011} (\bibinfo {year} {2002})},\ \Eprint
  {http://arxiv.org/abs/hep-ph/0109274} {arXiv:hep-ph/0109274 [hep-ph]}
  \BibitemShut {NoStop}%
\bibitem [{\citenamefont {Savage}(1997)}]{Savage:1996tb}%
  \BibitemOpen
  \bibfield  {author} {\bibinfo {author} {\bibfnamefont {M.~J.}\ \bibnamefont
  {Savage}},\ }\href {\doibase 10.1103/PhysRevC.55.2185} {\bibfield  {journal}
  {\bibinfo  {journal} {Phys. Rev.}\ }\textbf {\bibinfo {volume} {C55}},\
  \bibinfo {pages} {2185} (\bibinfo {year} {1997})},\ \Eprint
  {http://arxiv.org/abs/nucl-th/9611022} {arXiv:nucl-th/9611022 [nucl-th]}
  \BibitemShut {NoStop}%
\bibitem [{\citenamefont {Pastore}\ \emph {et~al.}(2009)\citenamefont
  {Pastore}, \citenamefont {Girlanda}, \citenamefont {Schiavilla},
  \citenamefont {Viviani},\ and\ \citenamefont {Wiringa}}]{Pastore:2009is}%
  \BibitemOpen
  \bibfield  {author} {\bibinfo {author} {\bibfnamefont {S.}~\bibnamefont
  {Pastore}}, \bibinfo {author} {\bibfnamefont {L.}~\bibnamefont {Girlanda}},
  \bibinfo {author} {\bibfnamefont {R.}~\bibnamefont {Schiavilla}}, \bibinfo
  {author} {\bibfnamefont {M.}~\bibnamefont {Viviani}}, \ and\ \bibinfo
  {author} {\bibfnamefont {R.~B.}\ \bibnamefont {Wiringa}},\ }\href {\doibase
  10.1103/PhysRevC.80.034004} {\bibfield  {journal} {\bibinfo  {journal} {Phys.
  Rev.}\ }\textbf {\bibinfo {volume} {C80}},\ \bibinfo {pages} {034004}
  (\bibinfo {year} {2009})},\ \Eprint {http://arxiv.org/abs/0906.1800}
  {arXiv:0906.1800 [nucl-th]} \BibitemShut {NoStop}%
\bibitem [{\citenamefont {Detmold}\ and\ \citenamefont
  {Savage}(2004)}]{Detmold:2004qn}%
  \BibitemOpen
  \bibfield  {author} {\bibinfo {author} {\bibfnamefont {W.}~\bibnamefont
  {Detmold}}\ and\ \bibinfo {author} {\bibfnamefont {M.~J.}\ \bibnamefont
  {Savage}},\ }\href {\doibase 10.1016/j.nuclphysa.2004.07.007} {\bibfield
  {journal} {\bibinfo  {journal} {Nucl. Phys.}\ }\textbf {\bibinfo {volume}
  {A743}},\ \bibinfo {pages} {170} (\bibinfo {year} {2004})},\ \Eprint
  {http://arxiv.org/abs/hep-lat/0403005} {arXiv:hep-lat/0403005 [hep-lat]}
  \BibitemShut {NoStop}%
\bibitem [{\citenamefont {Chen}\ \emph
  {et~al.}(1999{\natexlab{b}})\citenamefont {Chen}, \citenamefont {Rupak},\
  and\ \citenamefont {Savage}}]{Chen:1999vd}%
  \BibitemOpen
  \bibfield  {author} {\bibinfo {author} {\bibfnamefont {J.-W.}\ \bibnamefont
  {Chen}}, \bibinfo {author} {\bibfnamefont {G.}~\bibnamefont {Rupak}}, \ and\
  \bibinfo {author} {\bibfnamefont {M.~J.}\ \bibnamefont {Savage}},\ }\href
  {\doibase 10.1016/S0370-2693(99)01007-2} {\bibfield  {journal} {\bibinfo
  {journal} {Phys.Lett.}\ }\textbf {\bibinfo {volume} {B464}},\ \bibinfo
  {pages} {1} (\bibinfo {year} {1999}{\natexlab{b}})},\ \Eprint
  {http://arxiv.org/abs/nucl-th/9905002} {arXiv:nucl-th/9905002 [nucl-th]}
  \BibitemShut {NoStop}%
\bibitem [{\citenamefont {Fleming}\ \emph {et~al.}(2000)\citenamefont
  {Fleming}, \citenamefont {Mehen},\ and\ \citenamefont
  {Stewart}}]{Fleming:1999ee}%
  \BibitemOpen
  \bibfield  {author} {\bibinfo {author} {\bibfnamefont {S.}~\bibnamefont
  {Fleming}}, \bibinfo {author} {\bibfnamefont {T.}~\bibnamefont {Mehen}}, \
  and\ \bibinfo {author} {\bibfnamefont {I.~W.}\ \bibnamefont {Stewart}},\
  }\href {\doibase 10.1016/S0375-9474(00)00221-9} {\bibfield  {journal}
  {\bibinfo  {journal} {Nucl.Phys.}\ }\textbf {\bibinfo {volume} {A677}},\
  \bibinfo {pages} {313} (\bibinfo {year} {2000})},\ \Eprint
  {http://arxiv.org/abs/nucl-th/9911001} {arXiv:nucl-th/9911001 [nucl-th]}
  \BibitemShut {NoStop}%
\bibitem [{\citenamefont {Lynn}\ \emph {et~al.}(2016)\citenamefont {Lynn},
  \citenamefont {Tews}, \citenamefont {Carlson}, \citenamefont {Gandolfi},
  \citenamefont {Gezerlis}, \citenamefont {Schmidt},\ and\ \citenamefont
  {Schwenk}}]{Lynn:2015jua}%
  \BibitemOpen
  \bibfield  {author} {\bibinfo {author} {\bibfnamefont {J.~E.}\ \bibnamefont
  {Lynn}}, \bibinfo {author} {\bibfnamefont {I.}~\bibnamefont {Tews}}, \bibinfo
  {author} {\bibfnamefont {J.}~\bibnamefont {Carlson}}, \bibinfo {author}
  {\bibfnamefont {S.}~\bibnamefont {Gandolfi}}, \bibinfo {author}
  {\bibfnamefont {A.}~\bibnamefont {Gezerlis}}, \bibinfo {author}
  {\bibfnamefont {K.~E.}\ \bibnamefont {Schmidt}}, \ and\ \bibinfo {author}
  {\bibfnamefont {A.}~\bibnamefont {Schwenk}},\ }\href {\doibase
  10.1103/PhysRevLett.116.062501} {\bibfield  {journal} {\bibinfo  {journal}
  {Phys. Rev. Lett.}\ }\textbf {\bibinfo {volume} {116}},\ \bibinfo {pages}
  {062501} (\bibinfo {year} {2016})},\ \Eprint
  {http://arxiv.org/abs/1509.03470} {arXiv:1509.03470 [nucl-th]} \BibitemShut
  {NoStop}%
\bibitem [{\citenamefont {Lynn}\ \emph {et~al.}(2017)\citenamefont {Lynn},
  \citenamefont {Tews}, \citenamefont {Carlson}, \citenamefont {Gandolfi},
  \citenamefont {Gezerlis}, \citenamefont {Schmidt},\ and\ \citenamefont
  {Schwenk}}]{Lynn:2017fxg}%
  \BibitemOpen
  \bibfield  {author} {\bibinfo {author} {\bibfnamefont {J.~E.}\ \bibnamefont
  {Lynn}}, \bibinfo {author} {\bibfnamefont {I.}~\bibnamefont {Tews}}, \bibinfo
  {author} {\bibfnamefont {J.}~\bibnamefont {Carlson}}, \bibinfo {author}
  {\bibfnamefont {S.}~\bibnamefont {Gandolfi}}, \bibinfo {author}
  {\bibfnamefont {A.}~\bibnamefont {Gezerlis}}, \bibinfo {author}
  {\bibfnamefont {K.~E.}\ \bibnamefont {Schmidt}}, \ and\ \bibinfo {author}
  {\bibfnamefont {A.}~\bibnamefont {Schwenk}},\ }\href {\doibase
  10.1103/PhysRevC.96.054007} {\bibfield  {journal} {\bibinfo  {journal} {Phys.
  Rev.}\ }\textbf {\bibinfo {volume} {C96}},\ \bibinfo {pages} {054007}
  (\bibinfo {year} {2017})},\ \Eprint {http://arxiv.org/abs/1706.07668}
  {arXiv:1706.07668 [nucl-th]} \BibitemShut {NoStop}%
\bibitem [{\citenamefont {Lonardoni}\ \emph {et~al.}(2018)\citenamefont
  {Lonardoni}, \citenamefont {Gandolfi}, \citenamefont {Lynn}, \citenamefont
  {Petrie}, \citenamefont {Carlson}, \citenamefont {Schmidt},\ and\
  \citenamefont {Schwenk}}]{Lonardoni:2018nob}%
  \BibitemOpen
  \bibfield  {author} {\bibinfo {author} {\bibfnamefont {D.}~\bibnamefont
  {Lonardoni}}, \bibinfo {author} {\bibfnamefont {S.}~\bibnamefont {Gandolfi}},
  \bibinfo {author} {\bibfnamefont {J.~E.}\ \bibnamefont {Lynn}}, \bibinfo
  {author} {\bibfnamefont {C.}~\bibnamefont {Petrie}}, \bibinfo {author}
  {\bibfnamefont {J.}~\bibnamefont {Carlson}}, \bibinfo {author} {\bibfnamefont
  {K.~E.}\ \bibnamefont {Schmidt}}, \ and\ \bibinfo {author} {\bibfnamefont
  {A.}~\bibnamefont {Schwenk}},\ }\href {\doibase 10.1103/PhysRevC.97.044318}
  {\bibfield  {journal} {\bibinfo  {journal} {Phys. Rev.}\ }\textbf {\bibinfo
  {volume} {C97}},\ \bibinfo {pages} {044318} (\bibinfo {year} {2018})},\
  \Eprint {http://arxiv.org/abs/1802.08932} {arXiv:1802.08932 [nucl-th]}
  \BibitemShut {NoStop}%
\bibitem [{\citenamefont {Butler}\ and\ \citenamefont
  {Chen}(2000)}]{Butler:1999sv}%
  \BibitemOpen
  \bibfield  {author} {\bibinfo {author} {\bibfnamefont {M.}~\bibnamefont
  {Butler}}\ and\ \bibinfo {author} {\bibfnamefont {J.-W.}\ \bibnamefont
  {Chen}},\ }\href {\doibase 10.1016/S0375-9474(99)00682-X} {\bibfield
  {journal} {\bibinfo  {journal} {Nucl. Phys.}\ }\textbf {\bibinfo {volume}
  {A675}},\ \bibinfo {pages} {575} (\bibinfo {year} {2000})},\ \Eprint
  {http://arxiv.org/abs/nucl-th/9905059} {arXiv:nucl-th/9905059 [nucl-th]}
  \BibitemShut {NoStop}%
\bibitem [{\citenamefont {Butler}\ \emph {et~al.}(2002)\citenamefont {Butler},
  \citenamefont {Chen},\ and\ \citenamefont {Vogel}}]{Butler:2002cw}%
  \BibitemOpen
  \bibfield  {author} {\bibinfo {author} {\bibfnamefont {M.}~\bibnamefont
  {Butler}}, \bibinfo {author} {\bibfnamefont {J.-W.}\ \bibnamefont {Chen}}, \
  and\ \bibinfo {author} {\bibfnamefont {P.}~\bibnamefont {Vogel}},\ }\href
  {\doibase 10.1016/S0370-2693(02)02868-X} {\bibfield  {journal} {\bibinfo
  {journal} {Phys. Lett.}\ }\textbf {\bibinfo {volume} {B549}},\ \bibinfo
  {pages} {26} (\bibinfo {year} {2002})},\ \Eprint
  {http://arxiv.org/abs/nucl-th/0206026} {arXiv:nucl-th/0206026 [nucl-th]}
  \BibitemShut {NoStop}%
\bibitem [{\citenamefont {Chen}\ \emph {et~al.}(2003)\citenamefont {Chen},
  \citenamefont {Heeger},\ and\ \citenamefont {Robertson}}]{Chen:2002pv}%
  \BibitemOpen
  \bibfield  {author} {\bibinfo {author} {\bibfnamefont {J.-W.}\ \bibnamefont
  {Chen}}, \bibinfo {author} {\bibfnamefont {K.~M.}\ \bibnamefont {Heeger}}, \
  and\ \bibinfo {author} {\bibfnamefont {R.~G.~H.}\ \bibnamefont {Robertson}},\
  }\href {\doibase 10.1103/PhysRevC.67.025801} {\bibfield  {journal} {\bibinfo
  {journal} {Phys. Rev.}\ }\textbf {\bibinfo {volume} {C67}},\ \bibinfo {pages}
  {025801} (\bibinfo {year} {2003})},\ \Eprint
  {http://arxiv.org/abs/nucl-th/0210073} {arXiv:nucl-th/0210073 [nucl-th]}
  \BibitemShut {NoStop}%
\bibitem [{\citenamefont {Ando}\ \emph {et~al.}(2008)\citenamefont {Ando},
  \citenamefont {Shin}, \citenamefont {Hyun}, \citenamefont {Hong},\ and\
  \citenamefont {Kubodera}}]{Ando:2008va}%
  \BibitemOpen
  \bibfield  {author} {\bibinfo {author} {\bibfnamefont {S.}~\bibnamefont
  {Ando}}, \bibinfo {author} {\bibfnamefont {J.~W.}\ \bibnamefont {Shin}},
  \bibinfo {author} {\bibfnamefont {C.~H.}\ \bibnamefont {Hyun}}, \bibinfo
  {author} {\bibfnamefont {S.~W.}\ \bibnamefont {Hong}}, \ and\ \bibinfo
  {author} {\bibfnamefont {K.}~\bibnamefont {Kubodera}},\ }\href {\doibase
  10.1016/j.physletb.2008.08.040} {\bibfield  {journal} {\bibinfo  {journal}
  {Phys. Lett.}\ }\textbf {\bibinfo {volume} {B668}},\ \bibinfo {pages} {187}
  (\bibinfo {year} {2008})},\ \Eprint {http://arxiv.org/abs/0801.4330}
  {arXiv:0801.4330 [nucl-th]} \BibitemShut {NoStop}%
\bibitem [{\citenamefont {De-Leon}\ \emph {et~al.}(2019)\citenamefont
  {De-Leon}, \citenamefont {Platter},\ and\ \citenamefont
  {Gazit}}]{De-Leon:2016wyu}%
  \BibitemOpen
  \bibfield  {author} {\bibinfo {author} {\bibfnamefont {H.}~\bibnamefont
  {De-Leon}}, \bibinfo {author} {\bibfnamefont {L.}~\bibnamefont {Platter}}, \
  and\ \bibinfo {author} {\bibfnamefont {D.}~\bibnamefont {Gazit}},\ }\href
  {\doibase 10.1103/PhysRevC.100.055502} {\bibfield  {journal} {\bibinfo
  {journal} {Phys. Rev.}\ }\textbf {\bibinfo {volume} {C100}},\ \bibinfo
  {pages} {055502} (\bibinfo {year} {2019})},\ \Eprint
  {http://arxiv.org/abs/1611.10004} {arXiv:1611.10004 [nucl-th]} \BibitemShut
  {NoStop}%
\bibitem [{\citenamefont {Acharya}\ and\ \citenamefont
  {Bacca}(2020)}]{Acharya:2019fij}%
  \BibitemOpen
  \bibfield  {author} {\bibinfo {author} {\bibfnamefont {B.}~\bibnamefont
  {Acharya}}\ and\ \bibinfo {author} {\bibfnamefont {S.}~\bibnamefont
  {Bacca}},\ }\href {\doibase 10.1103/PhysRevC.101.015505} {\bibfield
  {journal} {\bibinfo  {journal} {Phys. Rev.}\ }\textbf {\bibinfo {volume}
  {C101}},\ \bibinfo {pages} {015505} (\bibinfo {year} {2020})},\ \Eprint
  {http://arxiv.org/abs/1911.12659} {arXiv:1911.12659 [nucl-th]} \BibitemShut
  {NoStop}%
\bibitem [{\citenamefont {Pastore}\ \emph {et~al.}(2008)\citenamefont
  {Pastore}, \citenamefont {Schiavilla},\ and\ \citenamefont
  {Goity}}]{Pastore:2008ui}%
  \BibitemOpen
  \bibfield  {author} {\bibinfo {author} {\bibfnamefont {S.}~\bibnamefont
  {Pastore}}, \bibinfo {author} {\bibfnamefont {R.}~\bibnamefont {Schiavilla}},
  \ and\ \bibinfo {author} {\bibfnamefont {J.~L.}\ \bibnamefont {Goity}},\
  }\href {\doibase 10.1103/PhysRevC.78.064002} {\bibfield  {journal} {\bibinfo
  {journal} {Phys. Rev.}\ }\textbf {\bibinfo {volume} {C78}},\ \bibinfo {pages}
  {064002} (\bibinfo {year} {2008})},\ \Eprint {http://arxiv.org/abs/0810.1941}
  {arXiv:0810.1941 [nucl-th]} \BibitemShut {NoStop}%
\bibitem [{\citenamefont {Kolling}\ \emph {et~al.}(2011)\citenamefont
  {Kolling}, \citenamefont {Epelbaum}, \citenamefont {Krebs},\ and\
  \citenamefont {Mei{\ss}ner}}]{Kolling:2011mt}%
  \BibitemOpen
  \bibfield  {author} {\bibinfo {author} {\bibfnamefont {S.}~\bibnamefont
  {Kolling}}, \bibinfo {author} {\bibfnamefont {E.}~\bibnamefont {Epelbaum}},
  \bibinfo {author} {\bibfnamefont {H.}~\bibnamefont {Krebs}}, \ and\ \bibinfo
  {author} {\bibfnamefont {U.-G.}\ \bibnamefont {Mei{\ss}ner}},\ }\href
  {\doibase 10.1103/PhysRevC.84.054008} {\bibfield  {journal} {\bibinfo
  {journal} {Phys.Rev.}\ }\textbf {\bibinfo {volume} {C84}},\ \bibinfo {pages}
  {054008} (\bibinfo {year} {2011})},\ \Eprint {http://arxiv.org/abs/1107.0602}
  {arXiv:1107.0602 [nucl-th]} \BibitemShut {NoStop}%
\bibitem [{\citenamefont {Piarulli}\ \emph {et~al.}(2013)\citenamefont
  {Piarulli}, \citenamefont {Girlanda}, \citenamefont {Marcucci}, \citenamefont
  {Pastore}, \citenamefont {Schiavilla},\ and\ \citenamefont
  {Viviani}}]{Piarulli:2012bn}%
  \BibitemOpen
  \bibfield  {author} {\bibinfo {author} {\bibfnamefont {M.}~\bibnamefont
  {Piarulli}}, \bibinfo {author} {\bibfnamefont {L.}~\bibnamefont {Girlanda}},
  \bibinfo {author} {\bibfnamefont {L.~E.}\ \bibnamefont {Marcucci}}, \bibinfo
  {author} {\bibfnamefont {S.}~\bibnamefont {Pastore}}, \bibinfo {author}
  {\bibfnamefont {R.}~\bibnamefont {Schiavilla}}, \ and\ \bibinfo {author}
  {\bibfnamefont {M.}~\bibnamefont {Viviani}},\ }\href {\doibase
  10.1103/PhysRevC.87.014006} {\bibfield  {journal} {\bibinfo  {journal} {Phys.
  Rev.}\ }\textbf {\bibinfo {volume} {C87}},\ \bibinfo {pages} {014006}
  (\bibinfo {year} {2013})},\ \Eprint {http://arxiv.org/abs/1212.1105}
  {arXiv:1212.1105 [nucl-th]} \BibitemShut {NoStop}%
\end{thebibliography}
\end{document}